\documentclass[]{BioPhysMath}
\usepackage[OT4,T1]{fontenc}
\usepackage[utf8]{inputenc}
\usepackage{polski}
\usepackage{graphicx}

\usepackage{amsmath}
\usepackage{amssymb}

\usepackage[english,polish]{babel} 

\usepackage{color} 
\usepackage{tabularx}
\RequirePackage[numbers,sort&compress]{natbib}

\usepackage[colorlinks=true]{hyperref}
  \hypersetup{
    pdftitle={Template BioPhysMath}, 
    pdfauthor=Nowy Autor, 
    colorlinks,
    urlcolor=blue,
    filecolor=magenta,
    citecolor=green, 
    linkbordercolor={1 1 1}, 
    citebordercolor={1 1 1},  
    urlbordercolor={ 1 1 1}  
  } 

\newcommand{\orcid}[1]{\href{https://orcid.org/#1}{\includegraphics[scale=.05]{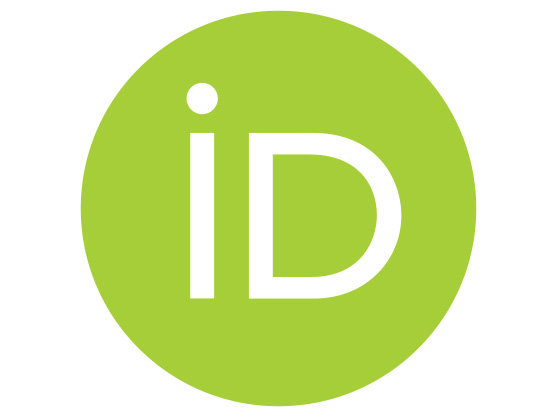}}}

\pdfoutput=1
\usepackage{graphicx}
\usepackage{wrapfig}
\graphicspath{{./Figures/},{./Pictures/}}
\usepackage{multicol}
\firstpage{i}    
\LogoG{\vspace*{1cm}\includegraphics[width=0.1\textwidth]{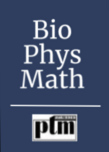}}
\volume{3}
\fasc{1}
\years{2024}
\LogoGMS{\includegraphics[width=0.18\textwidth]{bpm.png}}
\wwwtrue

\usepackage{makecell,longtable}
\setcellgapes{4pt} 
\usepackage{graphics,graphicx,subfigure}
\newcommand{\ben}{\begin{enumerate}}
\newcommand{\een}{\end{enumerate}}
\newcommand{\bit}{\begin{itemize}}
\newcommand{\eit}{\end{itemize}}
\newcommand{\be}{\begin{equation}}
\newcommand{\ee}{\end{equation}}
\newcommand{\bdm}{\begin{displaymath}}
\newcommand{\edm}{\end{displaymath}}
\newcommand{\bea}{\begin{eqnarray}}
\newcommand{\eea}{\end{eqnarray}}
\newcommand{\B}{\mathcal{B}}
\newcommand{\realnos}{\mathbb{R}}
\newcommand{\SM}{S}
\newcommand{\norm}[1]{\left\Vert#1\right\Vert}

\newtheorem{thm}{Theorem}[section]

\theoremstyle{plain}

\newtheorem{rem}{Remark}
\numberwithin{equation}{section}

\pages{38}
\received[20th June 2024]{28th November 2023}
\lastrevision{}
\logo{}{}{}

\title[Temporal Entropy Evolution]{Temporal Entropy Evolution in Stochastic and Delayed Dynamics}

\tpauthortrue 
\author[M.C. Mackey]{Michael C. Mackey \orcid{0000-0002-8524-2396}}
\affiliation{McGill University}
\address{Departments of Physiology, Physics \& Mathematics, McGill University, 3655 Promenade Sir William Osler, Montreal, Quebec H3G 1Y6, Canada}
\email{michael.mackey@mcgill.ca}
\city{Montreal}

\author[M. Tyran-Kamińska]{Marta Tyran-Kamińska\orcid{0000-0002-2630-792X}}
\affiliation{University of Silesia in Katowice}
\address{Institute of Mathematics,
University of Silesia in Katowice,
Bankowa 14,
40-007 Katowice,
Poland}
\email{mtyran@us.edu.pl}
\city{Katowice}

\comm{Krzysztof Burnecki}

\begin{document}
\vspace{-10ex}
\renewcommand{\thefootnote}{}
\footnote{\href{http://creativecommons.org/licenses/by/3.0/}{Licensed under a Creative Commons Attribution License (CC-BY)}}
\setcounter{page}{1} 
\selectlanguage{english}\Polskifalse

\begin{abstract} 
We review the behaviour of the Gibbs' and conditional entropies in deterministic and stochastic systems and continue to a formulation appropriate for a stochastically perturbed system with {\it delayed} dynamics.  The underlying question driving these investigations: ``Is the origin of the universally observed unidirectionality of time in our universe connected to the behaviour of entropy?''

We focus on temporal entropic behaviour with a review of previous results in deterministic and stochastic systems. Our emphasis is on the temporal behaviour of the Gibbs' and conditional entropies as these entropies give equilibrium results in concordance with experimental findings.  In invertible deterministic systems both entropies are temporally constant as has been well known for decades.  The addition of stochastic perturbations (a Wiener process) leads to an indeterminate (either increasing or decreasing) behaviour of the Gibbs' entropy, but the conditional entropy monotonically approaches equilibrium with increasing time.
The presence of delays in the dynamics, whether stochastically perturbed or not, leads to situations in which the Gibbs' and conditional entropies evolution can be oscillatory and not monotone, and may not approach equilibrium.
\end{abstract}

\section{Introduction}\label{s:intro}
\vskip 0.3cm

It is the universal experience that in all living things there is an inexorable one-way progression of events from birth through aging culminating in death.  We are faced daily with the inescapable conclusion that we will ultimately die.   We never witness the reverse sequence, nor will we ever.  Thus do our own lives and those of all living creatress highlight the apparent irreversibility in time of the universe as we perceive it.

The rather astonishing thing, however, is that all of the laws (evolution equations) that are written down in the physical sciences show no preference for a direction of time.  They are all equally valid for time $t$ going in a positive direction as well as in a negative direction, and this is true for the equations of mechanics, electricity and magnetism, special and general relativity and quantum mechanics (\citet{sachs87}) as well as in chemistry.  Why is there no clear preference for a direction of time in the dynamical equations of the physical sciences?  Much the same situation exists in biomathematics, but biomathematicians rarely if ever confront their own mortality from a research perspective.

Countless scientists have thought about, and written about, this apparently contradictory situation, the lack of temporal directionality in dynamical laws.  In a non-technical vein, \citet{brillouin1950thermodynamics} has written an extremely thoughtful essay examining a variety of issues related to this question that is informative,  deep,  and provocative. In his usual fashion,  Martin Gardner \cite{gardner1967can} has an amusing survey of the problems that would be encountered if time could go backward that is well worth reading and pondering.  Finally, \citet{gold1962arrow} expounds very eloquently on the possibility that temporal irreversibility is cosmologically derived from the (observed) expansion of the universe.  This is elaborated on in the essay by  \citet{schulman2010we}.

Many authors have considered the possible origins of the unidirectionality of time over the years, and without being exhaustive we mention \citet{reichenbach57}, \citet{fer1977irreversibilite},
\citet{davies1977physics}, \citet{sachs87},
\citet{zeh1992physical} and \citet{sklar93} as those  we have found worth reading and thought-provoking.  There is, in addition, a  very useful collection of reprinted musings on the subject in \citet{landsberg1985enigma}, as well as innumerable conference proceedings including
\citet{gold1969nature},
\citet{halliwell1996physical},
\citet{schulman97}, and
\citet{savitt1998time} to name but a few.

Not surprisingly, since many of those who have considered the possible origins of temporal unidirectionality have been physicists and chemists, the issue of the Second Law of Thermodynamics  has repeatedly been invoked, examined, and discussed.  In the early part of the 20$^{th}$ century Sir Arthur Eddington re-framed the issue of temporal unidirectionality in terms of the behaviour of  {\it entropy}, saying \cite{eddington2012new}:

\begin{quote}

``The law that entropy always increases holds, I think, the supreme position among the laws of Nature. If someone points out to you that your pet theory of the universe is in disagreement with Maxwell's equations - then so much the worse for Maxwell's equations. If it is found to be contradicted by observation - well, these experimentalists do bungle things sometimes. But if your theory is found to be against the Second Law of Thermodynamics I can give you no hope; there is nothing for it to collapse in deepest humiliation.''

\end{quote}

Here we have explored only a few of the possibilities for the uni-direction\-ality of time related to the temporal behaviour of entropy.  Our considerations are not new except for those of Section \ref{sec:dde}, but they do serve to highlight the nature of the problem.
In this review, we focus on the temporal behaviour of the Gibbs' and conditional entropies in dynamical and semi-dynamical systems with both stochastic perturbations as well as delayed dynamics, building on and extending the recent results of \citet{mackey2021can} as well as older results (\citet{mackey2006noise,mackey2006temporal}).

\renewcommand{\thefootnote}{\arabic{footnote}} 

The outline of the paper is as follows. Section \ref{sec:entropy} defines the the Gibbs' and conditional entropies, while Section \ref{s:det}
shows that these entropies are constant in time in invertible deterministic system (e.g. a system of ordinary
differential equations).
Section \ref{sec:AS}
introduces the dynamic concept of asymptotic stability,\footnotemark[1]\footnotetext[1]{Asymptotic stability is a strong
convergence property of ensembles which implies mixing.  Mixing, in
turn, implies ergodicity.}, and two main results
connecting the convergence of the conditional entropy with
asymptotic stability (Theorem~\ref{t:entropyconv}), and the
existence of unique stationary densities with the convergence of the
conditional entropy (Theorem~\ref{t:unique}).
Section \ref{s:gauss} considers the
stochastic extension where a system of ordinary differential equations is
perturbed by Gaussian white noise (thus becoming non-invertible) and
gives general results showing that in this stochastic case
asymptotic stability holds. In this section we give  examples of the effects of noise on the behaviour of the Gibbs and conditional entropy.
Following this we turn to an examination of the effects of delays in the dynamics on the evolution of the conditional entropy in Section \ref{sec:dde}.  We conclude with a short discussion, highlighting the confounding nature of our conclusions with respect to the role of delays, and suggesting further areas for study.

\section{Entropies: Gibbs and conditional}\label{sec:entropy}
\vskip 0.3cm

Two types of entropies are considered here.  The first is the {\it Gibbs' entropy}, which is
an extension of the equilibrium entropy
introduced by \citet{gibbs02} to a time dependent
situation.  This has been considered by a number of authors, namely \citet{ruelle96,ruelle03}, \citet{nicolis98},
\citet{nicolis99} and
\citet{bag02,bag02b,bag03}.

In
his seminal work \citet{gibbs02},  assuming the
existence of a system steady state  density $f_*$ on the
phase space ${ X}$, introduced the concept of the index of
probability given by $(-\ln f_*(x) )$ where ``$\ln$'' denotes
the natural logarithm.  He then identified the entropy in a
steady state  situation with the average of the index of
probability
\begin{equation}
H_G(f_*) = - \int_{ X}  f_*(x) \ln f_*(x)\, dx.
\label{e-gibbs}
\end{equation}
This is called the {\it equilibrium} or {\it steady state Gibbs'
entropy}.

The Gibbs' equilibrium entropy definition eq.~\eqref{e-gibbs} has
repeatedly proven to yield correct results when applied to a variety
of equilibrium situations  (\citet{mayer1940statistical,reichlbk,ter1995elements,kittel2004elementary,Penrose2005,hill2013statistical}) and this is why it is the gold standard for
equilibrium computations in statistical mechanics and
thermodynamics. Thus it makes total sense to  identify the
equilibrium  Gibbs' entropy ${H_G(f_*)}$ with the equilibrium
thermodynamic entropy.

The question of how a time dependent non-equilibrium entropy should
be defined has interested  investigators for some time, and
specifically the question of whether the Gibbs' entropy $H_G(f)$ should
be extended to a time dependent situation has occupied many researchers. Various aspects of
this question have been considered
(\citet{nicolis99,ruelle96,ruelle97,ruelle03,ruelle04}) and if the definition of the steady state Gibbs' entropy is extended to time
dependent (non-equilibrium) situations then the {\it time
dependent Gibbs' entropy} of a density $f(x,t)$ is defined~by
\begin{equation}
H_G(f) = - \int_{ X}  f(x,t) \ln f(x,t)\, dx.
\label{e-gibbstime}
\end{equation}

A second type of entropy, the {\it conditional entropy}, is a generalization of the Gibbs' entropy.
Convergence properties of the conditional entropy have been
extensively studied because entropy methods  have been
known for some time to be useful for problems involving
questions related to convergence of solutions in partial
differential equations
(\citet{loskot91,abbond99,toscanivillani,arnold01,qian01,qian02}).
Explicitly, we define the {\it conditional entropy} as   (\citet{almcmbk94}) 
    \begin{equation}
H_c(f|f_*)=-\int_X f(x,t)\ln\dfrac{f(x,t)}{f_*(x)}dx.
\label{d:conent}
    \end{equation}
This is also known as
the Kullback-Leibler or relative entropy (\citet{loskot91}), and has been related to the thermodynamic free
energy (\citet{qian02a}).

Before proceeding mention should be made of the Boltzmann entropy, and there are several different versions all going by the same name (c.f.  \citet{uffink2022sep-statphys-Boltzmann} who has sorted this out very nicely).  We speak only about the original version of Boltzmann's entropy (\citet{Boltzmann1872})  considered by \citet{jaynes65}.

In spite of the fact that the Boltzmann entropy has played a prominent role in considerations of the arrow of time, is that it suffers from several severe flaws.  First it does not give correct equilibrium entropies in a number of physical situations whereas the Gibbs' entropy is routinely used for equilibrium calculations (\citet{jaynes1971violation}).  Secondly the Boltzmann and Gibbs' entropies agree only in the situation of a dilute gas in which there are no interactions between constituents (\citet{jaynes65}).  Finally there are clear examples in which the temporal behaviour of the Boltzmann entropy  shows non-physical behaviour but the Gibbs' does not (\citet{mackey2006temporal}).  In spite of these clear divergences, there are investigators (\citet{lebowitz93,lebowitz99,lebowitz99a}) for whom Boltzmann clarified all issues related to the directionality of time.  Much has been written on the subject of Gibbs' versus Boltzmann and for the interested reader we refer to \citet{uffink06}, \citet{frigg2016field} and \citet{frigg2019statistical}.

\pagebreak 

\section{Gibbs' and conditional entropy in invertible deterministic systems}
\label{s:det}
\vskip 0.3cm

To set the stage let $(X,\B,\mu)$ be a $\sigma$-finite measure space. Further, let
$\{P^t\}_{t\ge 0}$  be a semigroup of Markov operators on
$L^1(X,\mu)$, {\it i.e.} $P^tf\ge 0$ for $f\ge 0$, $\int P^tf(x)\,
\mu(dx)=\int f(x) \, \mu(dx)$, and $P^{t+s}f=P^t(P^sf)$. If the
group property holds for $t,s \in \realnos$, then we say that $P$ is
invertible, while if it holds only for $t,s \in \realnos ^+$ we say
that $P$ is non-invertible. We denote the corresponding set of
densities by $\mathcal{D}(X,\mu)$, or $\mathcal{D}$ when there will
be no ambiguity, so $f \in \mathcal{D}$ means $f \geq 0$ and
$||f||_1=\int_X f(x) \,\mu(dx) = 1$. A density $f_*$ is called a stationary density of $P^t$ if $P^tf_*=f_*$ for all $t$.

In this section we briefly consider entropy behaviour in situations where the dynamics are invertible in the
sense that they can be run forward or backward in time without ambiguity.
To
make this clearer, consider a phase space $X$ and a dynamics $S_t:X \to X$. For
every initial point $x^0$, the sequence of successive points $S_t(x^0)$,
considered as a function of time $t$, is called a {\bf trajectory}. In the
phase space $X$, if the trajectory $S_t(x^0)$ is nonintersecting with itself,
or intersecting but periodic, then at any given final time $t_f$ such that $x^f
= S_{t_f}(x^0)$ we could change the sign of time by replacing $t \rightarrow  (-t)$, and
run the trajectory backward using $x^f$ as a new initial point in $X$. Then our
new trajectory $S_{-t}(x^f)$ would arrive precisely back at $x^0$ after a time
$t_f$ had elapsed: $x^0 = S_{-t_f}(x^f)$.  Thus in this case we have a dynamics
that may be reversed in time {\it completely unambiguously}.

We formalize this by introducing the concept of a {\bf dynamical system} $\lbrace S_t \rbrace _{t \in
\realnos}$, which is simply any group of transformations $S_t:X \rightarrow X$ having the two properties: 1.
$S_0 (x) = x$; and 2. $S_t(S_{t'}(x)) = S_{t+t'}(x)$ for $t,t'\in \mathbb{R}  $ or $\mathbb{Z}$. Since, from the
definition, for any $t \in  \mathbb{R}$, we have $S_t(S_{-t}(x)) = x = S_{-t}(S_t(x))$, it is clear that dynamical
systems are invertible in the sense discussed above since they may be run either forward or backward in time.
Systems of ordinary differential equations are examples of dynamical systems.

The first result is very general, and shows that the conditional entropy of any
invertible system is constant and uniquely determined by the method of system
preparation.  Thus

\begin{thm}
\label{thm-invert}\cite[Thm. 3.2]{mackey2011time}
If $P^t $ is an invertible Markov operator and has a   stationary density
$f_*$, then the conditional entropy is constant  and equal to the value
determined by $f_*$ and the choice of the initial density $f_0$ for all time
$t$.  That is,
\begin{equation*}
H_c (P^tf_0 |f_*) \equiv H_c(f_0|f_*) 
\end{equation*}
for all $t$.
\end{thm}

In the case where we are considering a deterministic dynamics $S_t: {\cal X} \to {\cal X}$ where ${\cal X}
\subset X$, then the corresponding Markov operator is also known as the Frobenius Perron operator
(\citet{almcmbk94}), and is given explicitly~by
    \begin{equation}\nonumber
    P^tf_0(x) = f_0(S_{-t}(x)) | J^{-t}(x)|
    \end{equation}
where $J^{-t}(x)$ denotes the Jacobian of $S_{-t}(x)$. A simple calculation shows
    \begin{align}
    H_c(P^tf_0|f_*) &= -\int_ {\cal X} P^t f_0(x)\log \left [ \dfrac{P^t f_0(x)}{f_*(x)}\right ]\, dx \nonumber \\
    &= -\int_ {\cal X} f_0(S_{-t}(x))|J^{-t}(x)|\log \left [\dfrac{ f_0(S_{-t}(x)) }{f_*(S_{-t}(x))}\right ]\, dx \nonumber \\
    &= -\int_ {\cal X} f_0(y)\log\left [ \dfrac{ f_0( y) } {f_*(y)}\right ] \, dy \nonumber \\
    &\equiv  H_c(f_0|f_*) \nonumber
       \end{align}
as expected from Theorem \ref{thm-invert}.

 More specifically, if the dynamics corresponding to our invertible Markov operator are described by the
system of ordinary differential equations
    \begin{equation}
    \dfrac{dx_i}{dt} = F_i(x) \qquad        i = 1,\ldots ,d
    \label{ode}
    \end{equation}
operating in a region of $\realnos^d$ with initial conditions $x_i(0) = x_{i,0}$, then (\citet{almcmbk94}) the
evolution of $f(x,t) \equiv P^tf_0(x)$ is  governed by the generalized Liouville  equation
\begin{equation}
\frac {\partial f}{\partial t} = -\sum_i \frac {\partial (fF_i)}{\partial x_i}. \label{e-leqn}
\end{equation}
The corresponding stationary density  $f_*$ is given by the solution of
\begin{equation*}
\sum_i \frac {\partial (f_* F_i)}{\partial x_i} = 0. \end{equation*}
Note that the uniform density  $f_* \equiv 1$, meaning that the flow defined by eq. \eqref{ode} preserves the
Lebesque measure,  is a stationary density of eq. \eqref{e-leqn} if and only if
\begin{equation*}
\sum_i \frac {\partial F_i}{\partial x_i} = 0.
\end{equation*}

In particular, for the system of ordinary differential equations (\ref{ode}) whose density evolves according to
the Liouville equation (\ref{e-leqn}) we can assert that the conditional entropy of the density $P^tf_0$ with
respect to the stationary density $f_*$ will be constant for all time and will have the value determined by the
initial density $f_0$ with which the system is prepared. This result can also be proved directly by noting that
from the definition of the conditional entropy we may write
$$
H_c(f|f_*) = -  \int_{\realnos ^d}  f(x) \left[ \log \left( \frac{f}{f_*} \right) + \frac{f_*}{f} - 1 \right] \,
dx
$$
when the stationary density is $f_*$.  Differentiating with respect to time gives
$$
\frac {dH_c}{dt}   = - \int_{\realnos ^d}  \frac {df}{dt} \log \left[ \frac {f}{f_*} \right] \,dx
$$
or, after substituting from (\ref{e-leqn}) for $(\partial   f/\partial   t)$, and integrating by parts under the
assumption that $f$ has compact support,
$$
\frac {dH_c}{dt} = \int_{\realnos ^d}  \frac{f}{f_*} \sum_i \frac {\partial (f_*F_i)}{ \partial x_i} \,dx.
$$
However, since $f_*$ is a stationary density of $P^t$, it is clear from (\ref{e-leqn}) that
\begin{equation}
\frac {dH_c}{dt} = 0,
\label{e:cond1}
\end{equation}
and we conclude that the conditional entropy $H_c(P^tf_0|f_*)$ does not change from its initial value when the
dynamics evolve in this manner.  It is obvious that this conclusion also holds for the Gibbs' entropy.

 \section{Asymptotic stability and entropy behaviour}\label{sec:AS}
\vskip 0.3cm

We call a semigroup of Markov
operators $P^t$ on $L^1(X,\mu)$ {\it asymptotically stable} if there
is a density $f_*$ such that $P^tf_*=f_*$ for all $t>0$ and for all
densities $f$
$$
\lim_{t\to\infty}||P^tf-f_*||_1=0.
$$

\begin{thm}{\cite[Thm. 7.7]{mackey2011time}}\label{t:entropyconv}
Let  $P^t$ be a semigroup of Markov operators on $L^1(X,\mu)$ and
$f_*$ be a positive density. If
    \begin{equation}
        \lim_{t \to \infty} H_c(P^tf|f_*) = 0
    \label{c:entconv}
    \end{equation}
for a density $f$ then
    \begin{equation}
    \lim_{t \to \infty} ||P^t f - f_*||_{1}=0.
    \label{c:l1conv}
    \end{equation}
Conversely,  if $P^tf_*=f_*$ for all $t\ge 0$  and Condition
{\em\eqref{c:l1conv}} holds for all $f$ with $H_c(f|f_*)>-\infty$, then
Condition {\em\eqref{c:entconv}} holds as well.
\end{thm}
Thus from this theorem {\it asymptotic stability is
necessary and sufficient for the convergence of $H_c$ to zero}.
Our next result draws a connection between the existence of a unique
stationary density $f_*$ of $P^t$ and the convergence of the
conditional entropy $H_c$.
Recall that the semigroup $P^t$ on $L^1(X,\mu)$ is called \textit{continuous} if   $P^tf$ converges to $f$ as $t\downarrow 0$ for every  density $f$  and it is {\it partially
integral} if there exists a measurable function $k : X \times X \to
[0,\infty)$ and $t_0>0$ such that
   $$
   P^{t_0}f(x) \ge \int_X k(x, y)f(y)\,\mu(dy)
    $$
for every density $f$ and
$$
\int_X \int_X k(x, y)\,\mu(dy)\,\mu(dx)>0.
$$

\begin{thm}{\cite[Thm. 2]{pichorrudnicki00}}\label{t:unique}
Let $P^t$ be a continuous partially integral semigroup of Markov operators. If
there is a unique stationary density $f_*$ for $P^t$ and $f_*>0$,
then $P^t$ is  asymptotically stable. In particular,
$$\lim_{t \to \infty} H_c(P^tf_0|f_*) = 0$$  for all $f_0$ with
$H_c(f_0|f_*)>-\infty$.
\end{thm}

 Since the
conditional and Gibbs' entropies are related by \[ H_G(P^tf_0
)=H_c(P^tf_0 |f_*)-\int P^tf_0(x)\log f_*(x) \mu(dx),
\]
 Theorem~\ref{t:entropyconv} implies the next result, see \citet{mackey2006temporal}.
\begin{thm}\label{th-convg}
Let  $P^t$ be an asymptotically stable semigroup of Markov operators
on $L^1(X)$ with a stationary density $f_*$ such that
$\displaystyle\int f_*^{1+r}(x)\, dx <\infty$ for all $r$ in some
neighborhood of zero. Then
$$
\lim_{t\to\infty}H_G(P^t f_0)=H_G(f_*)
$$
for all $f_0$ with $H_c(f_0|f_*)>-\infty$.
\end{thm}

These results are very general in their statements about the
behavior of the conditional and Gibbs' entropies. Theorem
\ref{thm-invert} tells us that when the dynamics are such that $P^t$
is a group (we have a dynamical system) the conditional entropy will
be constant and fixed by the initial value of $f_0$.  However, when
$P^t$ is an asymptotically stable semigroup then {\it Theorems
\ref{t:entropyconv} and \ref{th-convg} respectively guarantee the
convergence of the conditional entropy to its maximal value of zero
and  the Gibbs' entropy to its equilibrium value.}

\section{Effects of Gaussian noise}\label{s:gauss}
\vskip 0.3cm

In thermodynamic considerations physicists often assume and discuss an idealized situation of a system {\it isolated} from the environment,
and by that they mean that the system under consideration can exchange neither energy nor matter with its surroundings.   This is, of course, an unacheivable idealization.  In reality any system is subjected to perturbations from multiple outside sources, and from the Central Limit Theorem one would expect that the summated perturbations would be at least approximately Gaussian distributed.  Here we consider the effects of such a Gaussian distributed disturbance on dynamics and on entropies.

We consider  the behaviour of the stochastically perturbed system
    \begin{equation}
    \dfrac{dx_i}{dt} = F_i(x) + \sum_{j=1}^n \sigma_{ij}(x) \xi _j, \qquad   i = 1,\ldots ,d
    \label{stochode}
      \end{equation}
with the initial conditions $x_i(0) = x_{i,0}$,  where   $\sigma _{ij}(x)$ is the amplitude  of the stochastic
perturbation and  $\xi_j = \dfrac {dw_j}{dt}$ is a ``white noise'' term that is the derivative of a Wiener
process $w$. In matrix notation we can rewrite eq. \eqref{stochode} as
    \begin{equation}
dx(t)=F(x(t))dt + \sigma(x(t)) \, dw(t), \label{stochodem}
    \end{equation}
where $\sigma(x)=[\sigma_{ij}(x)]$ and $w$ is an $n$-dimensional Wiener process.\footnote{Here it is always assumed that the It\^o, rather that the
Stratonovich, calculus, is used.  For a discussion of the differences see \citet{VanKampen1981}, \citet{horsthemke84}, \citet{risken84}, and \citet{almcmbk94}. In particular, if the $\sigma_{ij}$ are independent of $x$ then the It\^o and  the
Stratonovich approaches yield identical results.}

The {\it Fokker-Planck equation}  that governs the evolution of the density function $f(x,t)$ of the process
$x(t)$ generated by the solution to the stochastic differential equation (\ref{stochodem}) is given by
(\citet{gardinerhandbook,risken84,van1992stochastic})
    \begin{equation}
    \frac {\partial f(x,t)}{\partial t} = -  \sum_{i=1}^d \frac{\partial
    [F_i(x)f(x,t)]}{\partial x_i} +\frac 12 \sum_{i,j=1}^d \frac{\partial ^2
    [a_{ij}(x)f(x,t)]}{\partial x_i \partial x_j}
    \label{fpeqn}
    \end{equation}
where
$$
a_{ij}(x)=\sum_{k=1}^{n}\sigma_{ik}(x)\sigma_{jk}(x).
$$

If $p(x,t|x_0)$ is the probability density function of $x(t)$ conditional on $x(0)=x_0$ then $p(x,t|x_0)$ is the fundamental solution of the Fokker-Planck equation, i.e.
for every $x_0$ the function $(x,t)\mapsto p(x,t|x_0)$ is a solution of the
Fokker-Planck equation with the initial condition $\delta(x-x_0)$. The
general solution $f(x,t)$ of the Fokker-Planck equation (\ref{fpeqn}) with the
initial condition
$$
f(x,0)=f_0(x)
$$
is then given by \begin{equation} f(x,t)=\int p(x,t|x_0)f_0(x_0)\, dx_0.
\label{gensoln}
\end{equation}
Define the Markov operators $P^t$ by 
    \begin{equation}
    P^tf_0(x)=\int p(x,t|x_0)f_0(x_0)\, dx_0, \quad f_0\in L^1.
    \label{mo}
    \end{equation}
Then $P^tf_0$ is the density of the solution $x(t)$ of eq. \eqref{stochodem} provided that $f_0$ is the density of
$x(0)$.

If the coefficients $a_{ij}$ and $F_i$ are sufficiently regular then
a fundamental solution $p$ exists.
One such set of  conditions is the
following: (1) the $F_i$ are of class $C^2$ with bounded derivatives; (2) the
$a_{ij}$ are of class $C^3$ and bounded with all derivatives bounded; and (3)
the uniform parabolicity condition holds, {\it i.e.} there exists a strictly
positive constant $\rho>0$ such that
$$
\sum_{i,j=1}^da_{ij}(x)\lambda_i\lambda_j\ge \rho \sum_{i=1}^d \lambda_i^2,\qquad \lambda_i,\lambda_j\in \realnos,
x\in \realnos^d.
$$
The uniform parabolicity condition implies that $p(t,x|x_0)>0$, and thus $P^tf(x)>0$ for every density, which
implies that there can be at most one stationary density, and that  if it exists then $f_*>0$.
The steady state density  $f_*(x)$ is the stationary solution of the Fokker Planck eq. (\eqref{fpeqn}):
    \begin{equation}
    -  \sum_{i=1}^d \frac{\partial
    [F_i(x)f]}{\partial x_i} +\frac 12 \sum_{i,j=1}^d \frac{\partial ^2
    [a_{ij}(x)f]}{\partial x_i \partial x_j} = 0.
    \label{ssfpeqn}
        \end{equation}
In this setting, asymptotic stability holds
 if and only if there is a stationary density $f_*$  that satisfies (\ref{ssfpeqn}).

{\bf Gibbs' entropy}

Differentiating eq.
\eqref{e-gibbstime} for the Gibbs' entropy yields
    \[
    \dfrac{dH_{G}}{dt} = \int  f \left(\sum_i \frac
    {\partial F_i}{\partial x_i} - \dfrac12  \sum_{i,j} \frac
    {\partial^2 a_{ij}(x)}{\partial x_i \partial x_j}\right) \,dx +
    \dfrac12 \int \dfrac{1}{f}
    \sum_{i,j=1}^d a_{ij}\dfrac
    {\partial f}{\partial x_i} \dfrac
    {\partial f}{\partial x_j}\, dx.
       \label{bgnoise}
    \]
If the $a_{ij}$ are independent of $x$ then we obtain
    \begin{equation}\nonumber
    \dfrac{dH_{G}}{dt} = \int  f \sum_i \frac
    {\partial F_i(x)}{\partial x_i}  \,dx +
    \dfrac12 \sum_{i,j=1}^d a_{ij}\int \dfrac{1}{f}
    \dfrac
    {\partial f}{\partial x_i} \dfrac
    {\partial f}{\partial x_j}\, dx.
       \label{c-bgnoise}
    \end{equation}
The first term
is of indeterminant sign (\citet[eq. (14)]{nicolis99}),  but the second is positive definite so
the {\it temporal behavior of the Gibbs' entropy in the presence of Gaussian noise is
unclear.}

{\bf Conditional entropy}

Differentiating eq.~\eqref{d:conent} for the conditional entropy with respect to time, and using
eq.~\eqref{fpeqn} with integration by parts, along with the fact that
since $f_*$ is a stationary density it satisfies (\ref{ssfpeqn}), we
obtain
    \begin{equation}
    \dfrac{dH_c}{dt} = \dfrac12 \int \left( \dfrac{f_*^2}{f} \right)
    \sum_{i,j=1}^{d} a_{ij}(x) \dfrac {\partial}{\partial x_i} \left( \dfrac {f}{f_*} \right)
    \dfrac {\partial}{\partial x_j} \left( \dfrac{f}{f_*} \right) \,dx.
    \label{dtcondent}
    \end{equation}
    Since the matrix $(a_{ij}(x))$ is nonnegative definite, one
concludes that
\begin{equation}
\dfrac{dH_c}{dt}\ge 0.
\label{e:cond2}
\end{equation}
This implies that the conditional entropy is a monotonic function of time.

We now discuss the speed of convergence to zero of the conditional entropy.
It is known that under some  conditions (\citet{bakryemery85,arnold01}) there exists a constant $\lambda>0$ such
    \[
    H_c(P^tf_0|f_*)\ge e^{-2\lambda t}H_c(f_0|f_*) 
    \]
for all initial densities $f_0$ with $H_c(f_0|f_*)>-\infty$. See \citet{mackey2006noise} for a review of these conditions.
Here we use a different approach.

Note that for $P^t$ as in \eqref{mo} we have the following lower bound for the conditional entropy
\begin{equation}\label{LB}
H_c(P^tf_0|f_*)\ge \int H_c(p(\cdot,t|x_0)|f_*)f_0(x_0)dx_0
\end{equation}
for any initial density $f_0$. To see this use Jensen's inequality to obtain
\[
\eta\left(\frac{P^tf_0(x)}{f_*(x)}\right)=\eta\left(\int \frac{p(x,t|x_0)}{f_*(x)}f_0(x_0)dx_0\right)\ge \int \eta\left(\frac{p(x,t|x_0)}{f_*(x)}\right)f_0(x_0)dx_0,
\]
where $\eta$ is the concave function  $\eta(s)=-s\log s$ for $s>0$. Next multiply by $f_*(x)$,  integrate with respect to $x$ and change the order of integration to obtain  the lower bound.

\subsection{Example of an Ornstein-Uhlenbeck process}\label{sss:1dou}
\vskip 0.3cm

Consider the scalar linear differential equation  with additive noise
\begin{equation}\label{1dOU}
dx(t)=ax(t)dt+\sigma dw(t),
\end{equation}
where $a,\sigma\in \mathbb{R}$ and $\sigma>0$.
If we take $a=-\gamma$ with $\gamma>0$ then eq. \eqref{1dOU} defines the Ornstein-Uhlenbeck process,  which was historically developed in thinking about perturbations to the velocity of a Brownian
particle.
The solution $x(t)$  of \eqref{1dOU} is of the form
\[
x(t)=e^{ta}x(0)+\sigma\int_0^te^{(t-r)a}dw(r).
\]
Note that
\[
\eta(t)=\sigma\int_0^te^{(t-r)a}dw(r),\quad t\ge 0,
\]
is a Gaussian process with mean zero and covariance of the form
\[
\mathbb{E}(\eta(t)\eta(t+s))=\sigma^2\int_0^t e^{(t-r)a}e^{(t+s-r)a}dr=\sigma^2 \int_0^t e^{ra}e^{(r+s)a}dr,\quad t,s\ge 0,
\]
by the It\^o isometry. If $x(0)=x_0$ then the random variable $x(t)$ has a Gaussian distribution with mean $\mathbb{E}(x(t))=e^{ta}x_0$ and variance
\[
\textrm{Var}(x(t))=\mathbb{E}([x(t)-\mathbb{E}(x(t))]^2)=\mathbb{E}(\eta(t)^2)=\sigma^2\int_0^t[e^{ra}]^2dr.
\]
Thus the probability density function of $x(t)$ conditional on $x(0)=x_0$ is given~by
    \begin{equation}
    p(x,t|x_0)=\frac{1}{\sqrt{2 \pi \upsilon(t)}}\exp \left \{-\frac{(x-e^{
    ta}x_0)^2}{2\upsilon(t)}\right \},
    \label{o-utdep}
    \end{equation}
where
\[
\upsilon(t):=\sigma^2\int_0^t[e^{ra}]^2dr. 
\]
If $x(0)$ has a Gaussian distribution with mean $0$ and variance $\sigma_*^2$ then $x(t)$ is again Gaussian with mean $0$ and variance
\[
\sigma_t^2=e^{2at}\sigma_*^2+\upsilon(t).
\]
Note that $\sigma_t^2$ is independent of $t$ if and only if
\begin{equation}\label{exOU}
a<0\quad \text{and}\quad \sigma_*^2=-\frac{\sigma^2}{2a}.
\end{equation}
If condition \eqref{exOU} holds then we define the process
\[
Y(t)=\sigma\int_{-\infty}^{t}e^{(t-r)a}dw(r),\quad t\in \mathbb{R},
\]
where $w$ is extended to $\mathbb{R}$ by taking an independent copy $\{\overline{w}(t):t\ge 0\}$ of the Wiener process and setting $w(t)=-\overline{w}(-t)$ for $t<0$. Observe that
\[
Y(t)=e^{ta}Y(0)+\eta(t),\quad t\ge 0.
\]
Thus $Y(t)$ is a solution of \eqref{1dOU} and $Y(t)$  has a Gaussian distribution with mean 0 and variance $\sigma_*^2$.
Consequently, $Y(t)$, $t\ge 0$,  is a stationary solution of \eqref{1dOU}. Its covariance function is of the form
\[
\mathbb{E}(Y(s)Y(t+s))=\sigma_*^2e^{ta}, \quad t\ge 0.
\]
Thus, if $a<0$ then
\[
\sigma_*^2=\lim_{t\to\infty}\sigma^2\int_0^t[e^{ra}]^2dr=\sigma^2\int_0^\infty[e^{ra}]^2dr<\infty
\] is the only solution of
\begin{equation}\label{e:ibv0}
2a\sigma_*^2+\sigma^2=0.
\end{equation}
We have
\[
\lim_{t\to \infty}p(x,t|x_0)=f_*(x),
\]
where
\[
f_*(x)=\frac{1}{\sqrt{2 \pi \sigma_*^2}}\exp \left \{-\frac{x^2}{2\sigma_*^2}\right \}
\]
is the density of $Y(t)$.
The function $f_*$ is then the unique stationary solution to
 the corresponding Fokker-Planck equation \begin{equation}\nonumber
    \dfrac{\partial f}{\partial t}
    =  -\dfrac {\partial [a x f]}{\partial x}
    + \dfrac {\sigma^2}2 \dfrac {\partial^2 f}{\partial x^2}.
    \label{o-ufpeqn}
    \end{equation}

We may now examine how the two different types of entropy behave.

{\bf Gibbs' entropy}

It is easily seen that the Gibbs' entropy of the Gaussian density
\[
f_{m,\sigma^2}(x)=\frac{1}{\sqrt{2\pi \sigma^2}}\exp{\left\{-\frac{1}{2\sigma^2}(x-m)^2\right\}}, \quad x\in \mathbb{R},
\] where $\sigma>0$ and $m\in \mathbb{R}$, is
\begin{equation}
H_G(f_{m,\sigma^2}) = -\int _{-\infty}^{+\infty}f_{m,\sigma^2}(x) \ln f_{m,\sigma^2}(x) dx = \dfrac 12 + \dfrac 12\ln (2 \pi \sigma^2).
\label{e:gibbs-1d}
\end{equation}

It is not the case that the Gibbs' entropy will always be an increasing function of time when the dynamics are perturbed by an Ornstein-Uhlenbeck process. To see this consider as the initial density $f_0$ a Gaussian
    \begin{equation}
    f_0(x) = \dfrac {1}{ \sqrt{2 \pi\bar \sigma^2}} \exp{\left \{ -\dfrac {x^2}{2 \bar \sigma^2} \right
    \}}\nonumber
    \end{equation}
where $\bar \sigma >0$, then
    \begin{equation}
    P^tf_0 (x) = \dfrac {1}{ \sqrt{2 \pi\bar{\sigma}_t^2}} \exp \left
    \{ - \dfrac {x ^2}{2 \bar{\sigma}_t^2} \right \}\nonumber
    \end{equation}
where
    \begin{equation}
    \bar{\sigma}^2_t = \sigma_*^2 + (\bar \sigma^2 -
    \sigma_*^2)e^{2a t}\nonumber
    \end{equation}
with $ \sigma_*^2 = -\sigma^2/2a$ and $a<0$.
  The Gibbs' entropy is
 \begin{equation}
    H_{G}(P^tf_0 ) = \dfrac 12+ \dfrac 12\ln (2 \pi\bar{\sigma}_t^2) \nonumber
    \end{equation}
and
   \begin{equation} \label{e:gibbs} \dfrac {dH_{G}(P^tf_0 )}{dt}  \left \{
    \begin{array}{lll}
    &>& 0  \qquad \mbox{for} \qquad \bar \sigma^2 < \sigma_*^2 \\
    &=& 0  \qquad \mbox{for}  \qquad \bar \sigma^2 = \sigma_*^2\\
    &<& 0   \qquad \mbox{for}  \qquad \bar \sigma^2 > \sigma_*^2
    ,
    \end{array} \right.  \end{equation}
 implying that the evolution  of the Gibbs' entropy in time is
a function of the statistical properties ($\bar \sigma ^2$)
of the initial ensemble.

Note also that the Gibbs' entropy of the transition probability $p$ is an increasing function of time, since
\[
H_{G}(p(\cdot,t|x_0)) =  \dfrac 12 +\dfrac 12\ln (2 \pi \upsilon(t))
\]
and the variance $\upsilon(t)$ is a strictly increasing function.

{\bf Conditional entropy}

Next, let us calculate the conditional entropy of two Gaussian
densities $f_{m_1,\sigma_1^2}$ and $f_{m_2,\sigma_2^2}$, where $\sigma_1,\sigma_2>0$ and  $m_1,m_2\in \realnos$.
We have
$$
\ln \dfrac{f_{m_1,\sigma_1^2}(x)}{f_{m_2,\sigma_2^2}(x)}=\ln \sqrt{\dfrac{\sigma_2^2}{\sigma_1^2}}-\frac{1}{2\sigma_1^2}(x-m_1)^2+\frac{1}{2\sigma_2^2}(x-m_2)^2.
$$
Since
$$
\int_{\realnos} f_{m_1,\sigma_1^2}(x)x^2\, dx=\sigma_1^2+m_1^2\quad\mbox{and}\quad \int_{\realnos}f_{m_1,\sigma_1^2}(x)x\,
dx=m_1,
$$
we arrive at
    \begin{equation}
H_c(f_{m_1,\sigma_1^2}|f_{m_2,\sigma_2^2})=\dfrac 12 \ln
\dfrac{\sigma_1^2}{\sigma_2^2}+\dfrac{1}{2}\left(1-\dfrac{\sigma_1^2}{\sigma_2^2}\right)-\dfrac{1}{2\sigma_2^2}(m_1-m_2)^2. \label{e:conent1dg}
    \end{equation}

To obtain the speed of convergence of the conditional entropy to zero we use the lower bound from~\eqref{LB}.
Observe that if we calculate the conditional entropy for the transition density $p$ as in \eqref{o-utdep} then
\begin{equation}\label{ent:OU1d}
H_c(p(\cdot,t|x_0)|f_*)=\frac12\ln \frac{\upsilon(t)}{\sigma_*^2}+\frac12\left(1-\frac{\upsilon(t)}{\sigma_*^2}\right)-\frac{1}{2\sigma_*^2}(e^{ta }x_0)^2,
\end{equation}
where $\sigma_*^2=\lim_{t\to\infty} \upsilon(t)$.
Consequently, we have for the one dimensional linear SDE with additive noise
\[
H_c(P^tf_0|f_*)\ge \frac12\ln \frac{\upsilon(t)}{\sigma_*^2}+\frac12\left(1-\frac{\upsilon(t)}{\sigma_*^2}\right)-\frac{1}{2\sigma_*^2}\int_{\mathbb{R}} (e^{ta }x_0)^2 f_0(x_0)dx_0.
\]
Since
\[
\lim_{t\to\infty} \upsilon(t)=\sigma_*^2\quad \text{and}\quad \lim_{t\to\infty}e^{ta }x_0=0
\]
we obtain $\lim_{t \to \infty}H_c(P^tf_0|f_*) = 0$ in addition to the conclusion in \eqref{e:cond2} that $\dot H_c(P^tf_0|f_*) \geq 0$.

\pagebreak

\section{The effects of delays}\label{sec:dde}
\vskip 0.3cm
\subsection{A Fokker-Planck-like formulation}\label{sec:f-p form}
\vskip 0.3cm

We start with a few preliminaries.
Throughout, $C=C([-\tau,0],\mathbb{R}^d)$ denotes the Banach space of all continuous
functions $\phi\colon[-\tau,0]\to\mathbb{R}^d$ equipped with the supremum norm and the Borel $\sigma$-algebra.
We initially focus on   the equation without noise
\begin{equation}\label{e:dde}
\begin{split}
   dx(t)=&\mathcal{F}(x(t),x(t-\tau))dt, \quad t\ge 0, \\
    x(t)=&\phi(t),\quad t\in[-\tau,0],
  \end{split}
\end{equation}
where the initial function $\phi\in C$ and $\mathcal{F}$ are such that for each $\phi\in C$  there exists a continuous function
$x\colon[-\tau,\infty)\to \mathbb{R}^d$ so \eqref{e:dde} has a unique
global solution that depends continuously on~$\phi$. For each $t\ge 0$ define the solution map $\SM_t\colon
C\to C$ by
\begin{equation}\label{d:sol}
\SM_t(\phi)(s)=x_t(s)=x(t+s)\quad \text{for }
s\in[-\tau,0],
\end{equation}
where $x(t)$ is a solution of \eqref{e:dde} with $x_0=\phi$.
The transformation $(t,\phi)\mapsto \SM_t(\phi)$ is
continuous and $\{\SM_t\}_{t\in\mathbb{R}^+}$ is a semi-dynamical system on
$C$ (\citet{hale-lunel}).

Further, let $B(C)$ be the space of bounded Borel measurable functions $\psi \colon C\to \mathbb{R}$
with the supremum norm
\[
\|\psi \|_{\infty}=\sup_{\phi\in C}|\psi (\phi)|,
\]
where $|\cdot|$ is a norm on $\mathbb{R}^d$,
and  $\mathcal{M}(C)$ (resp. $\mathcal{M}_1(C)$) denote the space of finite (resp. probability) Borel measures on $C$.
For any $\psi \in B(C)$ and $\mu\in \mathcal{M}(C)$ we use the customary scalar product notation
\[
\langle \psi ,\mu\rangle=\int_C \psi (\phi)\mu(d\phi).
\]
$C_b\subset B(C)$ denotes the Banach space of all bounded uniformly continuous
functions $\psi\colon C\to \mathbb{R}$ with the supremum norm. Following \citet{dynkin1965markov} and \citet{mohammed84}, $\psi_t\in C_b$, $t>0$ is said to \emph{converge weakly} to $\psi\in C_b$  as $t\to 0^{+}$ (denoted by
$\mathrm{w-}\lim_{t\to 0}\psi_t=\psi$) if
\[
\lim_{t\to 0}\langle \psi_t,\mu\rangle=\langle \psi,\mu\rangle\quad \text{for each }
\mu\in \mathcal{M}(C).
\]
This is equivalent to $\lim_{t\to
0}\psi_t(\phi)=\psi(\phi)$ for every $\phi\in C$,  and
$\sup_t\norm{\psi_t}_{\infty}<\infty$.

A semigroup $\{T^t\}_{t\ge
0}$ of linear operators on the space $C_b$ is \emph{weakly continuous at} $0$ if $\mathrm{w-}\lim_{t\to 0}T^t\psi=\psi$ for all $\psi\in
C_b.$
The \emph{weak generator} $\mathcal{L}\colon \mathcal{D}(\mathcal{L})\subset
C_b\to C_b$ of the semigroup $\{T^t\}_{t\ge 0}$ is defined by (\citet{dynkin1965markov,mohammed84}) \begin{align*}
\mathcal{D}(\mathcal{L})&=\{\psi\in C_b: \mathrm{w-}\lim_{t\to
0}\frac{1}{t}\bigl(T^t\psi-\psi\bigr)\; \text{exists}\},\\
\mathcal{L}\psi &=\mathrm{w-}\lim_{t\to 0}\frac{1}{t}\bigl(T^t\psi-\psi\bigr).
\end{align*}
For $\psi \in \mathcal{D}(\mathcal{L})$ and $\mu \in \mathcal{M}(C)$ we have
\begin{equation}\label{e:weak}
\langle T^t \psi,\mu\rangle=\langle  \psi,\mu\rangle +\int_0^t \langle T^r(\mathcal{L}\psi),\mu\rangle  dr,\quad t>0.
\end{equation}
We next look at the version of \eqref{e:dde} with a perturbation:
\begin{equation}\label{e:sdde}
\begin{split}
   dx(t)=&\mathcal{F}(x(t),x(t-\tau))dt+\sigma(x(t),x(t-\tau))dw(t), \quad t\ge 0, \\
    x(t)=&\phi(t),\quad t\in[-\tau,0],
  \end{split}
\end{equation}
where $\{w(t)\}_{t\ge 0}$ is a stochastic perturbation, assumed to be a  Wiener process, with an amplitude $\sigma$, potentially dependent on $x(t)$ as well as $x(t-\tau)$. Assume that a pathwise solution  $x(t)$ of \eqref{e:sdde} exists, and let
\[
x_t(s)=x(t+s), \quad s\in [-\tau,0], t\ge 0.
\]
It is known (\citet[Chapter III]{mohammed84}) that $\{x_t\}_{t\ge 0}$ is a $C$-valued Markov process with transition semigroup
\[
T^t\psi(\phi)=\mathbb{E}(\psi(x_t)|x_0=\phi).
\]
The semigroup $\{T^t\}_{t\ge 0}$ is weakly continuous.
Let $\mu_0\in \mathcal{M}_1(C)$, and for each $t\ge 0$ define  the probability measure $\mu_t$ on the space $C$ as the distribution of $x_t$, so
\[
\langle \psi,\mu_t\rangle =\langle T^t \psi,\mu_0\rangle,\quad t\ge 0, \psi\in C_b.
\]
It thus follows from \eqref{e:weak} and a change of variables that
\begin{equation}\label{e:weakf}
\langle \psi,\mu_t\rangle=\langle  \psi,\mu_0\rangle +\int_0^t \langle \mathcal{L} \psi,\mu_r\rangle  dr
\end{equation}
for all $ t>0$ and $ \psi \in \mathcal{D}(\mathcal{L})$.
However, it is difficult   (see \citet[Chapter IV]{mohammed84}) to identify the domain $ \mathcal{D}(\mathcal{L})$ of the weak generator $ \mathcal{L}$, and so  we introduce an \emph{extended generator} for the process $x_t$. This is defined as  a linear operator $\mathcal{L}$
from its domain $\mathcal{D}$ to the set of all Borel measurable functions on $C$, where $\psi \in \mathcal{D}$ if  for each $t>0$ we have
\begin{equation*}
\int_0^t \langle |\mathcal{L} \psi|,\mu_r\rangle  dr <\infty
\end{equation*}
and \eqref{e:weakf} holds. 
Then $\{\mu_t\}_{t\ge 0}$ is the solution of the equation
\begin{equation*}\label{eq:infdim}
\frac{\partial}{\partial t}\mu_t=\mathcal{L}^*\mu_t.
\end{equation*}
We will use below a subset
of $\mathcal{D}$ that will allow us to change \eqref{eq:infdim} into a partial differential
equation \eqref{e:weakL}, see \citet{Huang} for the general study of path-distribution dependent stochastic differential equations.

Let $C_c^2(\mathbb{R}^d)$ denote the space of twice  continuously differentiable functions with compact support, and consider the differential operator from the space $C_c^2(\mathbb{R}^d)$ to the set of all Borel measurable functions on $C$  defined by
\[
Lg(\phi)=\sum_{i=1}^d \mathcal{F}_i(\phi(0),\phi(-\tau))\frac{\partial g}{\partial x_i} (\phi(0))+\frac12\sum_{i,j=1}^d a_{ij}(\phi(0),\phi(-\tau))\frac{\partial^2 g}{\partial x_i\partial x_j} (\phi(0)),
\]
for $\phi\in C$, $g\in C_c^2(\mathbb{R}^d).$
Here, $a$ is the matrix $\sigma\sigma^T$, where $\sigma^T$ is the transpose of the matrix $\sigma$.
Let $\mu(t)$ be the marginal distribution of the measure $\mu_t$, defined on $\mathbb{R}^d$  by
\[
\mu(t)(B)=\mu_t\{\phi\in C: \phi(0)\in B\}, \quad B\in \mathcal{B}(\mathbb{R}^d).
\]
It can be rewritten with the projection map $\pi_0\colon C\to \mathbb{R}^d$ defined by $\pi_0(\phi)=\phi(0)$, $\phi\in C$, as
 $\mu(t)=\mu_t\circ \pi_0^{-1}$. Thus $\mu(t)$ is the distribution of $x(t)$ for all $t\ge 0$.
Then  we say that $\{\mu_t\}_{t\ge 0}$ is a solution of the equation
\begin{equation*}
\frac{\partial}{\partial t}\mu(t)=L^*\mu_t
\end{equation*}
if for each $ t> 0$ and $ g\in C_c^2(\mathbb{R}^d)$ 
\begin{equation*}
\int_{0}^t \langle |Lg|,\mu_r\rangle dr<\infty
\end{equation*}
holds, and
\begin{equation}\label{e:gLd}
\int_{\mathbb{R}^d} g(x)\mu(t)(dx)=\int_{\mathbb{R}^d} g(x)\mu(0)(dx)+\int_{0}^t \langle Lg,\mu_r\rangle dr.
\end{equation}
Realize that this is equivalent to requiring $\psi\in \mathcal{D}$ and $\mathcal{L}\psi=Lg$ for $\psi=g\circ \pi_0$ and all $g\in C^2_c(\mathbb{R}^d)$. Furthermore, if we introduce the measure
\[
\nu(t)=\mu_t\circ \pi_{0,-\tau}^{-1},\quad t\ge 0,\] where $\pi_{0,-\tau}\colon C\to \mathbb{R}^d\times \mathbb{R}^d$ is the projection map $\pi_{0,-\tau}(\phi)=(\phi(0),\phi(-\tau))$, $\phi\in C$, then
\begin{multline*}
\langle Lg,\mu_t\rangle=\\\int_{\mathbb{R}^d}\int_{\mathbb{R}^d}\Big(\sum_{i=1}^d \mathcal{F}_i(x,y)\frac{\partial g}{\partial x_i} (x) +\frac12\sum_{i,j=1}^d a_{ij}(x,y)\frac{\partial^2 g}{\partial x_i\partial x_j}(x)\Big) \nu(t)(dx,dy),
\end{multline*}
by the change of variables formula. The measure $\nu(t)$ is the distribution of $(x(t),x(t-\tau))$.

Now suppose, additionally, that the measure $\nu(t)$ has a density $f_\nu$ with respect to the Lebesgue measure on $\mathbb{R}^{2d}$ i.e., $\nu(t)(dx,dy)=f_\nu(x, y,t)dxdy$. Then the measure $\mu(t)$ has a density $f$ with respect to the Lebesgue measure on $\mathbb{R}^d$ and
\[
f(x,t)=\int_{\mathbb{R}^d}f_\nu(x,y,t)dy.
\]
We also have
\[
f(y,t-\tau)=\int_{\mathbb{R}^d}f_\nu(x,y,t)dx, \quad t\ge \tau.
\]
Therefore, we can rewrite \eqref{e:gLd} as
\begin{multline*}
\int_{\mathbb{R}^d} g(x)f(x,t)dx=\int_{\mathbb{R}^d} g(x)f(x,0)dx
+\int_0^t \int_{\mathbb{R}^d}\int_{\mathbb{R}^d} \Big(\sum_{i=1}^d \mathcal{F}_i(x,y)\frac{\partial g}{\partial x_i} (x)\\+\frac12\sum_{i,j=1}^d a_{ij}(x,y)\frac{\partial^2 g}{\partial x_i\partial x_j}(x)\Big)f_\nu(x,y,r)dxdy dr,
\end{multline*}
which is the weak   form of the evolution equation
\begin{equation}\label{e:weakL}
\boxed{
\begin{aligned}
\frac{\partial}{\partial t} f(x,t)&=-\sum_{i=1}^d \frac{\partial}{\partial x_i}\int_{\mathbb{R}^d} \mathcal{F}_i(x,y)f_\nu(x,y,t)dy\\ 
&\quad
+\frac12\sum_{i,j=1}^d \frac{\partial^2}{\partial x_i\partial x_j}\int_{\mathbb{R}^d}a_{ij}(x,y)f_\nu(x,y,t)dy.
\end{aligned}
}
\end{equation}

Note that eq. \eqref{e:weakL}, {\bf {\it which is our main result of this section}},  reduces to the  Fokker-Planck equation if $\mathcal{F}$ and $a$ are independent of              $y$.  We will now rewrite it to deduce an  extension of eq. \eqref{dtcondent} for the temporal rate of change of the conditional entropy $H_c$.
Let us introduce the density of $x(t-\tau)$ conditional on $x(t)=x$
\begin{equation}\label{e:conden}
  f_\nu(y,t-\tau|x,t)=\frac{f_\nu(x,y,t)}{f(x,t)}.
\end{equation}
Then we can rewrite \eqref{e:weakL} in the form
\begin{equation}\label{e:weakLA}
\frac{\partial}{\partial t} f(x,t)=-\sum_{i=1}^d \frac{\partial}{\partial x_i}(\overline{F}_i(x,t)f(x,t))
+\frac12\sum_{i,j=1}^d \frac{\partial^2}{\partial x_i\partial x_j}(\overline{a}_{ij}(x,t)f(x,t)),
\end{equation}
where
\begin{align*}
 \overline{F}_i(x,t)&= \int_{\mathbb{R}^d} \mathcal{F}_i(x,y)f_\nu(y,t-\tau|x,t)dy,\\
 \quad \overline{a}_{ij}(x,t)&=\int_{\mathbb{R}^d}a_{ij}(x,y)f_\nu(y,t-\tau|x,t)dy, \quad i,j=1,\ldots,d.
\end{align*}
If there exists a stationary distribution for our process $\{x_t\}_{t\ge 0}$ such that the distribution of $x(t)$ has the stationary  density $f_*$ and the distribution of $(x(t),x(t-\tau))$ has a density $f_{\nu*}$, then the density of $x(t-\tau)$ conditional on $x(t)=x$ will satisfy
\begin{equation*}
  f_{\nu*}(y|x)=\frac{f_{\nu*}(x,y)}{f_*(x)}.
\end{equation*}
Consequently, it follows from \eqref{e:weakL} that $f_*$ solves the following equation
\begin{equation}\label{e:weakSt}
0=-\sum_{i=1}^d \frac{\partial}{\partial x_i}(\overline{F}_i^*(x)f_*(x))
+\frac12\sum_{i,j=1}^d \frac{\partial^2}{\partial x_i\partial x_j}(\overline{a}_{ij}^*(x)f_*(x)),
\end{equation}
where now
\begin{align*}
\overline{F}_i^*(x)&= \int_{\mathbb{R}^d} \mathcal{F}_i(x,y)f_{\nu*}(y|x)dy,\quad \\ \overline{a}_{ij}^*(x)&=\int_{\mathbb{R}^d}a_{ij}(x,y)f_{\nu*}(y|x)dy, \quad i,j=1,\ldots,d.
\end{align*}
Using \eqref{e:weakLA} and integration by parts we arrive at
\begin{align*}
\frac {dH_c}{dt}   &= - \int_{\realnos ^d}  \frac {\partial f(x,t)}{\partial t} \log \left( \frac {f(x,t)}{f_*(x)} \right) dx\\
&=-\int_{\realnos ^d} \sum_{i=1}^d \overline{F}_i(x,t)f_*(x)\frac{\partial}{\partial x_i}\left( \frac {f(x,t)}{f_*(x)}\right)dx \\
&\quad+\frac12\sum_{i,j=1}^d \int_{\realnos ^d}\frac{\partial}{\partial x_j}(\overline{a}_{ij}(x,t)f(x,t))\frac {f_*(x)}{f(x,t)} \frac{\partial}{\partial x_i}\left( \frac {f(x,t)}{f_*(x)}\right)dx.
\end{align*}
Since
\[
\frac{\partial}{\partial x_j}(\overline{a}_{ij}f)\frac{f_*}{f}=\frac{\partial}{\partial x_j}(\overline{a}_{ij}f_*)+\overline{a}_{ij}\frac{f_*^2}{f}\frac{\partial}{\partial x_j}\left(\frac{f}{f_*}\right),
\]
we obtain
\begin{align*}
\frac {dH_c}{dt}& =-\int_{\realnos ^d} \sum_{i=1}^d \Big(\overline{F}_i(x,t)f_*(x)-\frac12\sum_{j=1}^d\frac{\partial}{\partial x_j}(\overline{a}_{ij}(x,t)f_*(x))\Big)\frac{\partial}{\partial x_i}\left( \frac {f(x,t)}{f_*(x)}\right)dx\\
&\quad+\frac12\sum_{i,j=1}^d \int_{\realnos ^d}\frac {f_*^2(x)}{f(x,t)} \overline{a}_{ij}(x,t)\frac{\partial}{\partial x_j}\left(\frac{f(x,t)}{f_*(x)} \right) \frac{\partial}{\partial x_i}\left( \frac {f(x,t)}{f_*(x)}\right)dx.\nonumber
\end{align*}
Now suppose that $a_{ij}$ does not depend on $y$. Then $\overline{a}_{ij}(x,t)=a_{ij}(x)=\overline{a}_{ij}^*(x)$ and by using \eqref{e:weakSt} we obtain
\begin{align*}
\frac {dH_c}{dt}& =\int_{\realnos ^d} \sum_{i=1}^d \frac{\partial}{\partial x_i}\left([\overline{F}_i(x,t)-\overline{F}_i^*(x)]f_*(x)\right)\frac {f(x,t)}{f_*(x)}dx\\
&\quad+\frac12\sum_{i,j=1}^d \int_{\realnos ^d}\frac {f_*^2(x)}{f(x,t)} {a}_{ij}(x)\frac{\partial}{\partial x_j}\left(\frac{f(x,t)}{f_*(x)} \right) \frac{\partial}{\partial x_i}\left( \frac {f(x,t)}{f_*(x)}\right)dx.
\end{align*}
Consequently,
the last term in the above equation, as in \eqref{dtcondent}, is nonnegative and  the first term is of indeterminant sign.
\begin{rem}
Take $d=1$ and assume that $a=\sigma^2$ does not depend on $y$ as in  \citet{guillouzic99}.
Then we can rewrite \eqref{e:weakL} as
\begin{multline}\label{e:1dim}
 \frac{\partial}{\partial t} f(x,t)=- \frac{\partial}{\partial x}\left(f(x,t)\int_{\mathbb{R}} \mathcal{F}(x,y) f_\nu(y,t-\tau|x,t)dy\right)
\\+\frac12\frac{\partial^2}{\partial x^2}\left(\sigma^2(x)f(x,t)\right).
\end{multline}
Thus  \eqref{e:conden} and \eqref{e:1dim} correspond to equations (8) and (6) in \citet{guillouzic99}, where
   they use the letter $p$ for densities, while their $f$ is our $\mathcal{F}$. They also restrict the values of $x$ to an interval $(a,b)$.

Note that eq. \eqref{e:weakL}, and consequently \eqref{e:1dim}, is not
closed, as it contains the density $f_\nu$ of the pair $(x(t),x(t-\tau))$, thus one needs to determine
an equation for $f_\nu$  as well. That leads to an infinite hierarchy of equations. We refer the reader to
\citet{loos2021stochastic} for a thorough review of various approaches to this problem.
\end{rem}

To obtain a lower bound on the conditional entropy for the case of a stochastic differential equation with delays, we make use of the same idea as in Section \ref{s:gauss}. Suppose that $p(x,t|\phi)$ is the probability density function of the solution $x(t)=x_t(0)$ of \eqref{e:sdde} with the initial condition  being  a deterministic function $\phi$.  By extending eq.~\eqref{gensoln} to this case, we define
\begin{equation}\label{pdfd}
f(x,t)=\int_C p(x,t|\phi)\mu_0(d\phi),
\end{equation}
where $\mu_0$ is the distribution of the initial condition $x_0=\phi$,  which is now an element of the function space $C$. Then we obtain the lower bound on the conditional entropy
\begin{equation}\label{Hcpdfd}
H_c(f(\cdot,t)|f_*)\ge \int_C H_c(p(\cdot,t|\phi)|f_*)\mu_0(d\phi)
\end{equation}
as in eq. \eqref{LB}.

\subsection{A linear first order  example}\label{ssec:linear first order}
\vskip 0.3cm

For general systems like \eqref{e:sdde} we have shown that the density evolution equation is of the form \eqref{e:weakL}.  However, the machinery to derive analytic solutions to such formulations is essentially non-existent (c.f \citet{loos2021stochastic,Holubec2022} for examples drawn from physics)  and thus we turn to linearizations about a steady state to gain insight into dynamic behaviours.

Thus we consider the linear scalar differential delay equation
\begin{equation}\label{e:lin}
\begin{split}
x'(t)&=ax(t)+bx(t-\tau),\quad t>0,\\
x(t)&=\phi(t),\quad t\in [-\tau,0],
\end{split}
\end{equation}
where $a,b$ are real constants
and $\phi \colon [-\tau,0]\to \mathbb{R}$ is a continuous function.
We can write the solution of \eqref{e:lin} as (see  \citet[Section 1.6]{hale-lunel})
\[
x(t)=X(t)\phi(0)+b\int_{-\tau}^0 X(t-r-\tau)\phi(r)dr,\quad t\ge 0,
\]
where $X(t)$ is the fundamental solution of  \eqref{e:lin}, which means that $X(t)$ satisfies \eqref{e:lin} with  $X(t)=0$ for $t<0$ and $X(0)=1$. We have
\begin{equation}\label{e:funs}
X(t)=\sum_{k=0}^{\lfloor t/\tau\rfloor} e^{a(t-k\tau)} \frac{b^k}{k!}(t-k\tau)^k,\quad t\ge 0,
\end{equation}
where $\lfloor s\rfloor=\max\{k\in \mathbb{Z}: k\le s\}$. Thus for the solution map $S_t$ in \eqref{d:sol} we obtain
\[
S_t\phi(s)=X(t+s)\phi(0)+b\int_{-\tau}^0 X(t+s-q-\tau)\phi(q)dq,\quad t+s\ge 0, s\in [-\tau,0],t\ge0,
\]
and $S_t\phi(s)=\phi(t+s)$ for $t+s< 0$, $s\in [-\tau,0]$, $t\ge 0$.

Now consider the extension of \eqref{e:lin} to a  linear stochastic differential delay equation with additive noise
\begin{equation}\label{e:lins}
dx(t)=(ax(t)+bx(t-\tau))dt+\sigma dw(t),\quad t\ge0,
\end{equation}
a delayed Ornstein-Uhlenbeck process,  and the initial condition
\[
  x(t)=\phi(t), \quad t\in [-\tau,0].
\]
Then
\[
x(t)=X(t)\phi(0)+b\int_{-\tau}^0 X(t-q-\tau)\phi(q)dq+\sigma\int_0^t X(t-q)dw(q),
\]
where $X$ is the fundamental solution \eqref{e:funs}.
Note that
\[
y(t)=\sigma\int_0^t X(t-q)dw(q),\quad t\ge 0,
\]
is a Gaussian process with mean zero and covariance of the form
\[
\mathbb{E}(y(t)y(t+s))=\sigma^2\int_0^t X(t-q)X(t+s-q)dq=\sigma^2 \int_0^t X(q)X(q+s)dq,\quad t,s\ge 0,
\]
by It\^o's isometry. We thus have the following representation of the process $x_t$
\begin{equation*}
x_t=S_t\phi +y_t,
\end{equation*}
where $S_t$ is the solution map of the deterministic part \eqref{e:lin} and $y_t$ is defined by
\[
y_t(s)=\left\{
         \begin{array}{ll}
           y(t+s), & \hbox{if } s+t\ge 0,\\
           0, & \hbox{if }s+t<0.
         \end{array}
       \right.
\]
Since
\[
x_t(s)=S_t\phi(s)+y_t(s), \quad s\in [-\tau,0],
\]
and $y_t(s)$  has mean zero, we obtain  $\mathbb{E}(x_t(s))=S_t\phi(s)$ and
\[
\mathrm{cov}(x_t(s_1),x_t(s_2))=\mathbb{E}(y_t(s_1)y_t(s_2))=\sigma^2 \int_0^{t+s_1} X(q)X(q+s_2-s_1)dq
\]
for $-\tau \le s_1\le s_2\le 0$ and $t+s_1>0$. Consequently, $x_t$ is a Gaussian process with mean $m_t\colon [-\tau,0]\to \mathbb{R}$ given by
\[m_t(s)=S_t\phi(s), \quad s\in [-\tau,0],\] and the covariance function $r_t\colon [-\tau,0]\times[-\tau,0]\to \mathbb{R}$ satisfies
\[
r_t(s_1,s_2)=\left\{
               \begin{array}{ll}
                 0, & t+\min\{s_1,s_2\}\le 0, \\
                \sigma^2 \int_0^{t+\min\{s_1,s_2\}} X(q)X(q+|s_2-s_1|)dq , & t+\min\{s_1,s_2\}> 0.
               \end{array}
             \right.
\]

{\bf Existence of densities for the linear first order example}

We check whether the random variable $x(t)$ and the vector $(x(t),x(t-\tau))^T$ have densities when the initial condition is $x(t)=\phi(t)$ for $t\in [-\tau,0]$.
The measure $\mu(t)$, the distribution  of $x(t)=x_t(0)$, is Gaussian and has mean $m_t(0)=S_t\phi(0)$ and variance
$\upsilon(t)=r_t(0,0)$.
The variance $\upsilon(t)= r_t(0,0)$ is always positive for $t>0$, since
\begin{equation}\label{eq:varp}
\upsilon(t)=\sigma^2\int_{0}^t X^2(s)ds\ge \sigma^2\int_{0}^{\min\{t,\tau\}} e^{as}ds>0.
\end{equation}
Hence  the density $p$ of $\mu(t)$ is
\begin{equation}\label{e:densf}
p(x,t)=\frac{1}{\sqrt{2\pi\upsilon(t)}}\exp \left\{-\frac{[x-S_t\phi(0)]^2}{2\upsilon(t)}\right\}.
\end{equation}
We introduce the conditional probability density of $x_t(0)=x(t)$ conditional on $x_0=\phi$ as
\[
p(x,t|\phi):=p(x,t).
\]
 Further,
the measure $\nu(t)$, which is the distribution of the vector $(x(t),x(t-\tau))^T=(x_t(0),x_t(-\tau))^T$, is a two-dimensional Gaussian with mean vector $(m_t(0),m_t(-\tau))^T=(S_t\phi(0),S_t\phi(-\tau))^T$ and  covariance matrix
\begin{equation}\label{e:Qt}
Q_t=\left(
  \begin{array}{cc}
    r_t(0,0) & r_t(0,-\tau) \\
    r_t(-\tau,0) & r_t(-\tau,-\tau) \\
  \end{array}
\right).
\end{equation}
Notice that $ r_t(-\tau,-\tau)=0$  and $r_t(-\tau,0)=0$ for $t\le \tau$. Hence, if $t\le \tau$ then $\det (Q_t)=0 $ and $\nu(t)$ does not have a density.

We now show that  $\nu(t)$ has a density for $t>\tau$.
To see this observe that
\[
r_t(-\tau,0)=\sigma^2\int_0^{t-\tau} X(s)X(s+\tau)ds,
\]
and thus
\[
r_t(-\tau,0)^2\le \sigma^4\int_0^{t-\tau} X^2(s)ds\int_0^{t-\tau}X^2(s+\tau)ds
\]
by the Cauchy-Schwartz inequality. Consequently, for $t>\tau$ we obtain
\[
r_t(-\tau,0)^2\le \upsilon(t-\tau)(\upsilon(t)-\upsilon(\tau))<\upsilon(t-\tau)\upsilon(t),
\]
implying that $\det (Q_t)> 0$.

Let $p_\nu$ be the density of the measure $\nu(t)$ for  $t>\tau$.
In this example the evolution equation \eqref{e:weakL} reduces to
\begin{equation}\label{e:linear-evol}
\frac{\partial}{\partial t} p(x,t)=-\frac{\partial}{\partial x}(axp(x,t))-\frac{\partial}{\partial x}\int_{\mathbb{R}} b y p_\nu(x,y,t)dy+\frac12\frac{\partial^2}{\partial x^2}(\sigma^2p(x,t)).
\end{equation}
Since $\nu(t)$ is a 2-dimensional Gaussian distribution, we can write
\begin{equation}\label{e:question}
p_\nu(x,y,t)=p(x,t)p_\nu(y,t-\tau|x,t)
\end{equation}
where $p_\nu(y,t-\tau|x,t)$, the density of $x(t-\tau)$ conditional on $x(t)=x$, is again Gaussian with mean
\begin{equation*}
m_{|}(x,t)=S_t\phi(-\tau)+\frac{r_t(-\tau,0)}{\upsilon(t)}[x-S_t\phi(0)].
\end{equation*}
Consequently \eqref{e:linear-evol} takes the form
\begin{align}\label{e:weakL1}
 \frac{\partial}{\partial t} p(x,t)=-\frac{\partial}{\partial x}([ax+b m_{|}(x,t)]p(x,t))+\frac12\frac{\partial^2}{\partial x^2}(\sigma^2p(x,t)).
\end{align}
It follows that $p$ satisfies eq. \eqref{e:weakL1} if and only if
\begin{equation}\label{e:sigr}
\frac{d}{dt}\upsilon(t)=2a\upsilon(t)+2br_t(-\tau,0)+\sigma^2,\quad t>\tau.
\end{equation}
From \eqref{e:lins} we see  that \eqref{e:sigr} always holds.

{\bf Existence of stationary solutions for the linear first order example}

Define
\[
\alpha_0=\max\{\mathrm{Re}\lambda: \lambda=a +be^{-\lambda \tau}\}.
\]
   For each $\alpha>\alpha_0$, there is a constant $c$ such that $|X(t)|\le c e^{\alpha t}$ for all $t>0$ \cite[Chap. 1, Thm, 5.2]{hale-lunel}. In particular, if $\alpha_0<0$ then $X(t)$ converges to zero exponentially rapidly.

\begin{figure}[htbp!]
\centering
\includegraphics{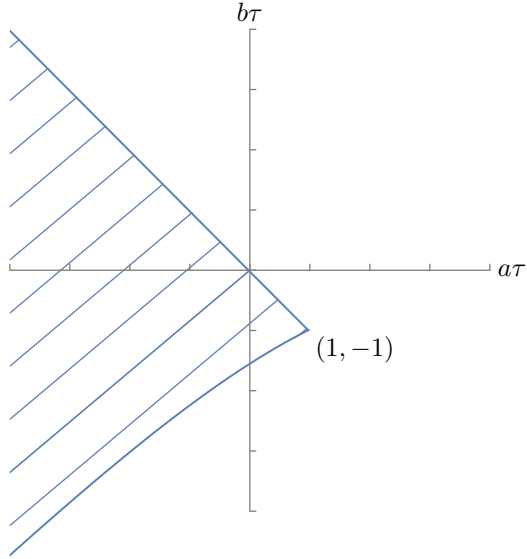}
\caption{A graphical representation of the stability of the steady state $x \equiv 0$ of \eqref{e:lin} as determined by \citet{hayes1950roots}.  All of the $(a,b)$ plane that is in the hatched wedge corresponds to parameter values $(a,b,\tau)$ such that $x= 0$ is stable.  For values of $(a,b,\tau)$ on the solid line boundary of the wedge there is an oscillatory periodic solution to \eqref{e:lin}.  Modified from \cite{mackey1995solution}.
}
\label{fig:stab}
\end{figure}

Based on the work of \citet{hayes1950roots}, see also \citet[Section 5.2 and Thm. A.5]{hale-lunel}, we have $\alpha_0<0$
if and only if
\begin{align}\label{e:alpha0}
a\tau <1,\quad  b\tau + a\tau<0,\quad  b\tau+a\tau \cos\kappa +\kappa\sin \kappa>0,
\end{align}
where $\kappa$ is the root of $\kappa=a\tau \tan \kappa$, $0<\kappa<\pi$ if $a\neq 0$ and $\kappa=\pi/2$ if $a=0$.
These are values  of $(a,b)$ inside the hatched wedge-shaped area of Figure \ref{fig:stab}.

In \citet[Proposition 2.8]{kuchler1992} it is shown that a stationary solution of \eqref{e:lins} exists if and only if $\alpha_0<0$ or, equivalently, the fundamental solution is square integrable:
\begin{equation}\label{e:ex}
\int_0^\infty X^2(q)dq<\infty.
\end{equation}
Then we have $\lim_{t\to\infty} S_t\phi(0)=0$ and
\begin{equation}\label{d:sigma}
\lim_{t\to\infty} \upsilon(t)=\lim_{t\to\infty}\sigma^2 \int_0^{t} X^2(q)dq=\sigma^2\int_0^\infty X^2(q)dq=:\sigma_*^2
\end{equation}
leading to
\[
\lim_{t\to\infty} p(x,t)=f_*(x)\quad \text{for all }x\in \mathbb{R},
\]
where
\begin{equation}\label{eq:stat den}
f_*(x)=\frac{1}{\sqrt{2\pi\sigma_*^2}}e^{-\frac{x^2}{2\sigma_*^2}}.
\end{equation}

If \eqref{e:ex} holds, then according to \citet{kuchler1992} a stationary solution $Z(t)$, $t\ge -\tau$, of \eqref{e:lins} can be represented as
\[
Z(t)=\sigma\int_{-\infty}^{t} X(t-s)dw(s),\quad t\in \mathbb{R},
\]
where $X$ is the fundamental solution \eqref{e:funs}, and $w$ is extended to $\mathbb{R}$. The covariance function of $\{Z(t): t\in \mathbb{R}\}$  is given by
\[
K(t)=\mathbb{E}(Z(s)Z(t+s))=\sigma^2\int_{0}^\infty X(q)X(q+t)dq,\quad t\ge 0,
\]
with $K(t):=K(-t)$ for $t<0$. We see that
\[
 K(0)=\sigma_*^2=\sigma^2 \int_0^\infty X^2(q)dq\quad \text{and}\quad K(\tau)=\lim_{t\to\infty}r_t(-\tau,0).
\]
Note that
\begin{equation}\label{e:ibv}
2a K(0)+2bK(\tau)+\sigma^2=0.
\end{equation}
To see this use the formula for $K(\tau)$ and  make use of \eqref{e:lin} for the fundamental solution $X$ to obtain
\[
bK(\tau)=\int_0^\infty b X(q-\tau)\sigma^2X(q)dq=\int_{0}^{\infty}(X'(q)-aX(q))\sigma^2X(q)dq,
\]
which gives
\[
bK(\tau)=\frac{\sigma^2}{2}\int_0^\infty \frac{d}{dq}[X^2(q)]dq -a K(0).
\]

Observe that if $\tau=0$ then condition \eqref{e:ibv} reduces to \eqref{e:ibv0}.  If $\tau>0$ then the value of  $\sigma^2_*=K(0)$ can be calculated as a function of the parameters $a,b,\sigma,\tau$ as in \citet[Proposition 2.13]{kuchler1992}, where it is shown that
\[
K(t)=K(0)g_1(t)+K'(0)g_2(t),\quad t\in [0,\tau],
\]
with
\[
g_1(t)=\left\{
         \begin{array}{ll}
           \cosh(lt), & \hbox{if $a^2>b^2$}, \\
           1, & \hbox{if $a^2=b^2$}, \\
           \cos(lt), & \hbox{if $a^2<b^2$,}
         \end{array}
       \right. \quad g_2(t)=\left\{
         \begin{array}{ll}
           \frac{1}{l}\sinh(lt), & \hbox{if $a^2>b^2$}, \\
           t, & \hbox{if $a^2=b^2$}, \\
           \frac{1}{l}\sin(lt), & \hbox{if $a^2<b^2$,}
         \end{array}
       \right.
\]
and
\begin{equation*}
l=\sqrt{|a^2-b^2|},\quad K(0)=\frac{\sigma^2}{2}\frac{bg_2(\tau)-1}{bg_1(\tau)+a},\quad  K'(0)=-\frac{\sigma^2}{2}. 
\end{equation*}

\begin{rem}
Observe that the covariance matrix $Q_t$ in \eqref{e:Qt} satisfies
\[
\lim_{t\to \infty}Q_t=\left(
                        \begin{array}{cc}
                          K(0) & K(\tau) \\
                          K(\tau) & K(0) \\
                        \end{array}
                      \right).
\]
Consequently, we have for the density $p_\nu$ as in \eqref{e:question}
\[
\lim_{t\to \infty}p_{\nu}(x,y,t)=f_*(x)f_*(y|x),\quad x,y\in \mathbb{R},
\]
where $f_*(y|x)$, the density of $Z(t-\tau)$ conditional on $Z(t)=x$, is of the form
\[
f_*(y|x)=\frac{1}{\sqrt{2\pi\sigma^2_*(1-\rho^2)}}
\exp\left\{-\frac{\left(y-\rho x\right)^2}{2\sigma^2_*(1-\rho^2)}\right\}\quad \text{with }\rho=\frac{K(\tau)}{K(0)} =-\frac{a}{b}-\frac{\sigma^2}{2b \sigma_*^2}.
\]
Note that eq. \ref{e:weakSt} for the stationary density $f_*$ is of the form
\[
0=-\frac{d}{dx}(\overline{F}^*(x)f_*(x))+\frac12\frac{d^2}{dx^2}(\sigma^2 f_*(x))
\]
with 
\[
\overline{F}^*(x)=\int(ax+by)f_*(y|x)dy=-\frac{\sigma^2}{2\sigma_*^2}x.
\]
\end{rem}

{\bf The Gibbs' and conditional entropy in the linear first order case}

Taking $f_{m_1,\sigma_1^2}$ to be given by \eqref{e:densf} and $f_{m_2,\sigma_2^2}$  by \eqref{eq:stat den},  we have $\sigma_1^2 = \upsilon(t), m_1 = m_t(0)=S_t\phi(0), \sigma_2^2 = \sigma^2_*, m_2 = 0$ and thus it follows from \eqref{e:gibbs-1d} that
\begin{equation*}
H_G(p(\cdot,t|\phi)) = \dfrac 12 + \dfrac 12 \ln (2 \pi \upsilon(t))
\end{equation*}
and, by \eqref{e:conent1dg} we have
\begin{equation}\label{dom initial condition}
H_c(p(\cdot,t|\phi)|f_*) = \dfrac 12 \ln \dfrac{\upsilon(t)}{\sigma^2_*} + \dfrac 12 \left ( 1 - \dfrac{\upsilon(t)}{\sigma^2_*} \right) - \dfrac{[S_t\phi(0)]^2}{2 \sigma^2_*}.
\end{equation}
Consequently, it follows from \eqref{Hcpdfd} that for the density $f$ of $x(t)$ with initial condition with distribution $\mu_0$ as defined in \eqref{pdfd} we obtain the following lower bound
\begin{equation}
H_c(f(\cdot,t)|f_*)\ge \frac12\ln \frac{\upsilon(t)}{\sigma_*^2}+\frac12\left(1-\frac{\upsilon(t)}{\sigma_*^2}\right)-\frac{1}{2\sigma_*^2}\int_C (S_t\phi(0))^2 \mu_0(d\phi).\label{e:ent-temp-inequald}
\end{equation}
Since $S_t\phi(0)\to 0$ as $t\to \infty$, we can conclude that
$H_c(f(t,\cdot)|f_*)\to 0$ for all initial distributions with finite second moment. Note that \eqref{e:ent-temp-inequald} gives an analogous lower bound for the entropy as in the non-delay case. We see that $S_t\phi(0)$ and $e^{ta}x_0$ are solutions of the corresponding deterministic differential equations, i.e. eqs. \eqref{e:lins} and \eqref{1dOU} with $\sigma=0$. %

\begin{figure}[htb]\centering
 \includegraphics[width=0.5\textwidth]{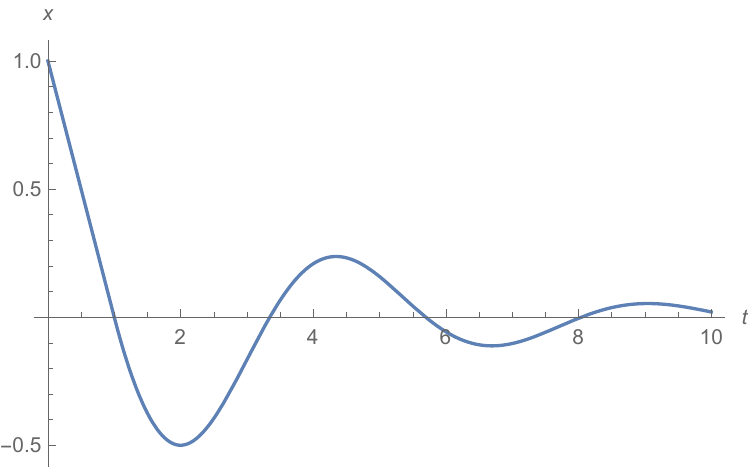}
\caption{The graph of a solution of equation $x'(t)=-x(t-1)$ with initial function $\phi\equiv 1$.  Note that this is for parameter values inside the hatched wedge of stability in Figure \ref{fig:stab}. }\label{f:damp}
\end{figure}

If we calculate the temporal rate of change then for the Gibbs' entropy
\begin{equation*}
\dfrac{dH_G(p(\cdot,t|\phi))}{dt} = \dfrac{1}{2 \upsilon(t)}\dfrac{d\upsilon(t)}{dt} \geq 0,
\end{equation*}
since $t\mapsto \upsilon(t)$ is an increasing and positive function of $t$ by \eqref{eq:varp}. This is precisely the same conclusion reached in \citet{mackey2006noise} for the Gibbs' entropy rate of change when the initial variance is zero.
For  the conditional entropy
$H_c$ we obtain
\begin{equation}
\dfrac{d H_c(p(\cdot,t|\phi)| f_*)}{dt}
= \dfrac{1}{2 \upsilon(t)}\dfrac{d \upsilon(t)}{dt} \left(1-\frac{\upsilon(t)}{\sigma_*^2}\right) -\frac{1}{2\sigma_*^2}\frac{d}{dt}[S_t\phi(0)]^2. \label{e:cond ent derivative}
\end{equation}
Now note that since $\upsilon(t) \leq \sigma^2_*$ by \eqref{d:sigma}, then for the first term in \eqref{e:cond ent derivative}
\[
\dfrac{1}{2 \upsilon(t)}\dfrac{d \upsilon(t)}{dt} \left(1-\frac{\upsilon(t)}{\sigma_*^2}\right) \geq 0.
\]
However we cannot assert that
\[
-\frac{1}{2\sigma_*^2}\frac{d}{dt}[S_t\phi(0)]^2 \geq 0,
\]
for the second term, and the
problem is that $S_t\phi(0)$ may exhibit a damped oscillation.  For example, consider $a=0$, $b=-1$ and $\tau=1.1$, and $\phi(s)=1$. Then $x'(t)=-1$ for $t\in[0,\tau]$, so  $S_t\phi(0)=x(t)=1-t$ for $t\in [0,\tau]$ and
\[
\frac{d}{dt}[S_t\phi(0)]^2=2x(t)x'(t)=-2(1-t)
\]
which  is either positive or negative for $t \in [0,1.1]$, see Figure \ref{f:damp}.

\begin{figure}[htb]\centering
\includegraphics[width=0.5\textwidth]{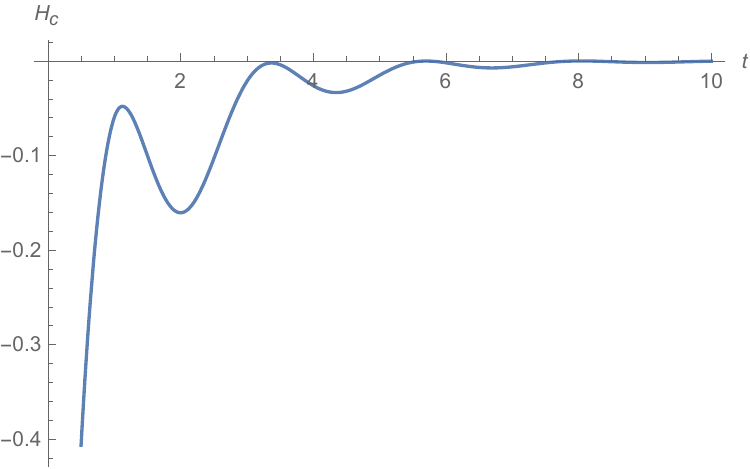}
\caption{The graph of the conditional entropy as given by \eqref{dom initial condition} for the example in Figure \ref{f:damp} and added noise with $\sigma=1/4
$.
}
\end{figure}

\pagebreak

{\bf Lack of monotonicity of the entropies in one dimensional linear delay dynamics with and without noise}

We now show that both the Gibbs' entropy $H_G(f(\cdot,t))$ and the conditional entropy $H_c(f(\cdot,t)|f_*)$ may not be monotone functions of time. To see this we take as the initial distribution a Gaussian distribution.
Suppose as in \citet{mackey2021can} that the initial condition $x_0=\phi$ is a Gaussian process $\xi$ with covariance function $R_0$ of the form
\begin{equation}\label{e:R0}
R_0(s_1,s_2)=\int_{-\tau}^{0}\eta_r(s_1)\eta_r(s_2)dr,
\end{equation}
for some  function $\eta_r\colon [-\tau,0]\to \mathbb{R}$ such that $(s,r)\mapsto \eta_r(s)$ is Borel measurable.
Then the covariance $R_t$ of the Gaussian process $S_t\xi$ is given by
\begin{equation*}
R_t(s_1,s_2)=\int_{-\tau}^0 S_t\eta_r(s_1)S_t \eta_r(s_2)dr
\end{equation*}
and in particular, for the variance $\sigma^2(t)=R_t(0,0)$  of $S_t\xi(0)$  we have
\begin{equation*}
\sigma^2(t)=\int_{-\tau}^0 (S_t\eta_r(0))^2dr.
\end{equation*}

One example of \eqref{e:R0} is
$R_0(s_1,s_2)=\bar\sigma^2(\min\{s_1,s_2\}+\tau)$ for $s_1,s_2\in [-\tau,0]$, where $\bar\sigma>0$, since  \eqref{e:R0} holds with
\begin{equation*}
\eta_r(s)=\bar\sigma1_{[r,0]}(s),\quad s,r\in [-\tau,0].
\end{equation*}
Then we have  $\xi(s)=\bar\sigma W(s+\tau)$, where $W=\{W(t)\}_{t\ge 0}$ is a standard Wiener process on $[0,\infty)$.
Consider now eq. \eqref{e:lins} with $a=0$
\begin{equation}\label{ex:lm}
dx(t)=bx(t-\tau)dt+\sigma dw(t), \quad t>0,
\end{equation}
and the initial condition
\[
x(t)=\xi(t), \quad t\in [-\tau,0],
\]
where we assume that $\xi$ is independent of the Wiener process $w$.
Then
\[
x(t)=S_t\xi(0)+y(t)
\]
has a Gaussian distribution with mean $0$ and  variance $\bar\sigma^2_t$ being the sum of variances of $S_t\xi(0)$ and $y(t)$
\[
\bar\sigma^2_t=\int_{-\tau}^0 (S_t\eta_r(0))^2dr+\upsilon(t).
\]
It is easily seen that
\[
S_t\eta_r(0)=\bar\sigma X(t-r), \quad t\ge 0, r\in[-\tau,0].
\]
Thus we obtain
\begin{equation}\label{sigD}
\sigma^2(t)=\bar\sigma^2\int_{t}^{t+\tau} X^2(q)dq,
\end{equation}
leading to
\begin{equation*}
\bar\sigma^2_t=\frac{\bar\sigma^2}{\sigma^2}\upsilon(t+\tau)+\left(1-\frac{\bar\sigma^2}{\sigma^2}\right)\upsilon(t).
\end{equation*}

\begin{figure}[htb]\centering
\includegraphics[width=0.5\textwidth]{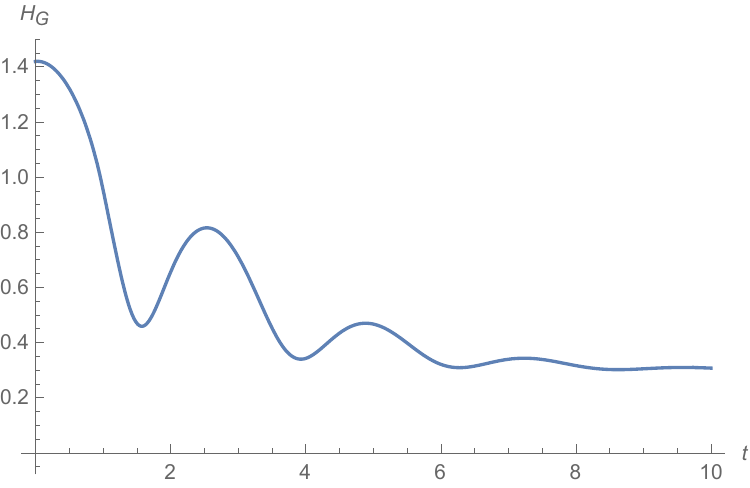}
\caption{The graph of the Gibbs' entropy  for the Gaussian density of the solution of \eqref{ex:lm} with $b=-1$, $\tau=\bar\sigma=1$ and $\sigma=1/4$}
\label{figGElm}
\end{figure}

\begin{figure}[htb]\centering
\includegraphics[width=0.5\textwidth]{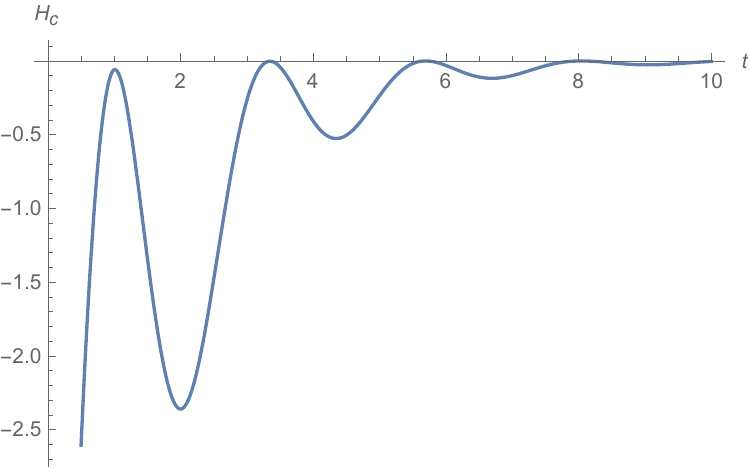}
\caption{The graph of the conditional entropy  for the Gaussian density of the solution of \eqref{ex:lm} with $b=-1$, $\tau=\bar\sigma=1$ and $\sigma=1/4$}
\label{figCElm}
\end{figure}

Note that if $\bar\sigma^2> \sigma^2$ then  $\bar\sigma^2_t$ need not be a monotone function of time, implying that the Gibbs's entropy and the conditional entropy of the density $f(x,t)$ of the solution of  \eqref{ex:lm} will not be monotone functions of time, see Figures \ref{figGElm} and \ref{figCElm}.


Finally, let us recall that eq. \eqref{ex:lm} has a stationary solution if and only if
\begin{equation}\label{climb}
-\frac{\pi}{2}<b\tau<0.
\end{equation} In that case we have $f(x,t)\to f_*(x)$ as $t\to \infty$ by \eqref{pdfd} and the fact that $p(x,t|\phi)\to f_*(x)$ as $t\to \infty$ for each $\phi\in C$ and arbitrary initial distribution.

\begin{figure}[htb]\centering
\includegraphics[width=0.5\textwidth]{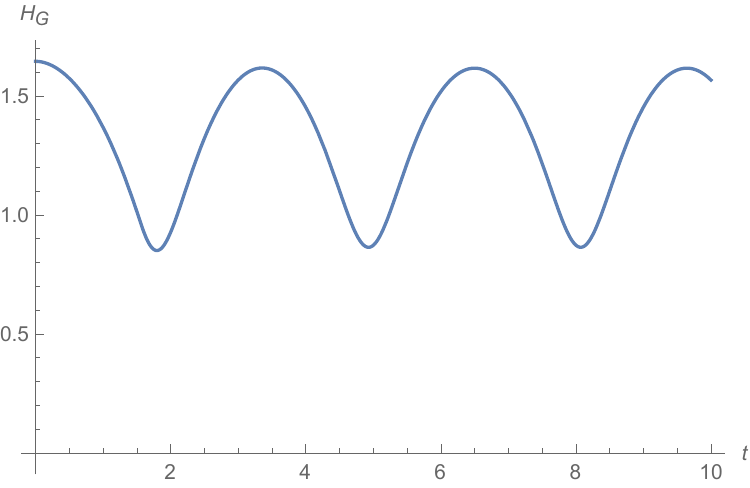}
\caption{The graph of the Gibbs' entropy  for the Gaussian density of the solution of $x'(t)=-x(t-\tau)$ with $\tau=\pi/2$, $\bar\sigma=1$ }
\label{figGElmd}
\end{figure}

Now,  suppose that $\sigma=0$ in  \eqref{ex:lm} and that we start with an initial distribution being Gaussian as above. If  condition \eqref{climb} holds then the variance $\sigma^2(t)$ of $x(t)$ as given by eq. \eqref{sigD} converges to $0$ as $t\to \infty$ and $0$ is a stationary solution of \eqref{ex:lm}, see \citet{mackey2021can}. If $b=-1$ and $\tau=\pi/2$, then  $x(t)$ has a Gaussian density $f(x,t)$ with variance $\sigma^2(t)$. The graph of its Gibbs' entropy is presented in Figure \ref{figGElmd}, showing clear sustained oscillations.

\begin{figure}[htb]\centering
\includegraphics[width=0.5\textwidth]{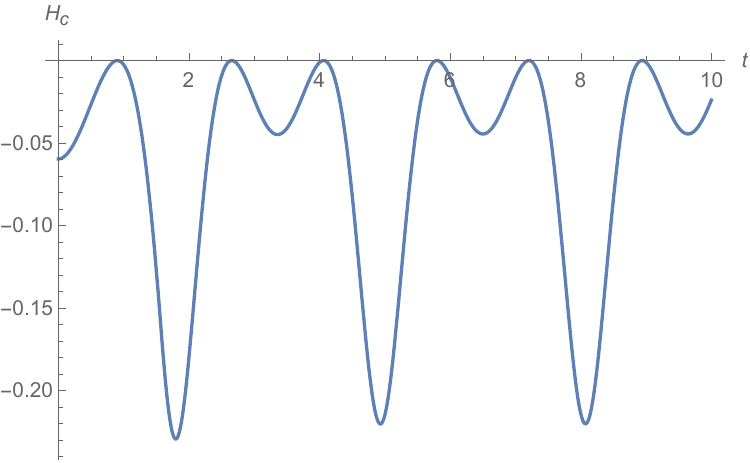}
\caption{The graph of the conditional entropy  for the Gaussian density of the solution of $x'(t)=-x(t-\tau)$ with $\tau=\pi/2$, $\bar\sigma=1$}
\label{figCElmd}
\end{figure}

 It is known (\citet{mackey2021can}) that equation $x'(t)=-x(t-\tau)$ with $\tau=\pi/2$ has a non-zero stationary solution in which case $f_*(x)$ is the Gaussian density with mean 0 and variance 1. Then the conditional entropy $H_c(f(\cdot,t)|f_*)$ is of the form
\[
H_c(f(\cdot,t)|f_*)=\frac12\ln \sigma^2(t)+\frac12(1-\sigma^2(t))
\]
and its graph is presented in Figure \ref{figCElmd}, again demonstrating the sustained and non-decrementing oscillations.


\section{Discussion}\label{sec:disc}
\vskip 0.3cm

Here we have reviewed and extended only a few of the possibilities for the uni-directionality of time related to the temporal behaviour of entropy.  Our considerations are not new except for the results of Section \ref{sec:dde}, but serve to highlight the nature of the problem.  In Table \ref{tab:1} we have summarized our results in terms of the temporal behaviours of $H_G$, $H_c$, and $H_{NE}$.
\begin{table}
\centering\makegapedcells
 \begin{tabular}{|p{3cm}||p{3cm}|p{3cm}|p{3cm}|}
   \hline
    Type of dynamics & $H_c(f|f_*)$ & $H_G(f)$ & $H_{NE}(f|f_*) = H_c(f|f_*) + H_G(f_*)$             \\ \hline \hline
   Invertible, Sec.~\ref{s:det}& $\dot H_c \equiv 0 $, Thm. \ref{thm-invert}, eq. \eqref{e:cond1} & $\dot H_G(f) \equiv 0$ & $\dot H_{NE}(f|f_*) \equiv 0$          \\ \hline
  Asymptotically  stable, Sec.~\ref{sec:AS} & $\lim_{t \to \infty} H_c(f|f_*) = 0$, Thm. \ref{t:entropyconv} & $\lim_{t \to \infty} H_G( f) = H_G(f_*)$, Thm. \ref{th-convg} &$\lim_{t\to\infty} H_{NE}(f|f_*) $ $ =H_G(f_*)$ \\ \hline
  Stochastic ODE, Sec.~\ref{s:gauss} & $\dot H_c \geq 0$, eq. \eqref{e:cond2} & $\dot H_G \,\, \pm$,  eq. \eqref{e:gibbs} & $\dot H_{NE} \geq 0 $ \\ \hline
  Delayed dynamics, Sec.~\ref{sec:dde}  & $\dot H_c \pm $, Fig. \ref{figCElmd} & $\dot H_G \pm $. Fig. \ref{figGElmd} & $\dot H_{NE} \pm $ \\ \hline
  Delayed stochastic, Sec.~\ref{sec:dde}  & $\dot H_c \pm $, Figs. \ref{f:damp}, \ref{figCElm} & $\dot H_G \pm $. Fig. \ref{figGElm} & $\dot H_{NE} \pm $ \\
    \hline
 \end{tabular}
 \caption{Summary of the behaviours of $H_c(P^tf|f_*)$, $H_G(P^tf)$, and $H_{NE}(P^tf|f_*)$ in which we indicate the sign of $\dot H_c$, $\dot H_{G}$, and $\dot H_{NE}$.}
 \label{tab:1}
\end{table}

The  non-equilibrium Gibbs' entropy $H_G(f)$ is manifestly not a
good candidate for $S_{TD}(t)$ because its dynamical behavior is at
odds with what is demanded by the Second Law of Thermodynamics.
As
we have demonstrated  in \citet{mackey2006temporal} and here, concrete analytic examples can
be constructed in which the direction of the temporal change in
$H_G(f)$ depends on the initial preparation of the system and others
can be constructed in which $H_G(f)$ oscillates in time. 

A number of other authors, including \citet[pp. 122-129, eq.
(247)]{degroot84},  \citet[pp. 111-114 and 185]{van1992stochastic}, and
\citet[p. 213]{Penrose2005}  have suggested that
a time dependent entropy should be associated dynamically with
      \begin{align}
    H_{NE}(f) &\equiv H_c(f|f_*e^{H_G(f_*)}) \nonumber \\
    &= H_c(f|f_*) + H_G(f_*)
    \end{align}
as an extension of the \citet[pp. 44-45 and 168]{gibbs02} discussion
of entropy.  This also goes under the name of the ``Gibbs' entropy
postulate'' (\citet{mazur94,mazur98a,mazur98b,mazur00,mazur01,rubi01}). 

In the absence of both stochastic perturbation and delays, we have shown that the conditional entropy (and thus $H_{NE}$) is temporally constant.  The introduction of stochasticity can induce the monotone approach of $H_{NE}$ to a maximum of zero.

In our quest to extend the problem of entropy evolution to situations with delays, in Section \ref{sec:dde} we have considered `density' evolution in stochastically perturbed systems with delayed dynamics like \eqref{e:sdde}.  We  have derived a `Fokker-Planck' like equation \eqref{e:weakL} for the `density'.  This is exactly analogous to our procedure in \citet{mackey2021can} when we derived a `Liouville-like' equation \cite[Eqn. 22]{mackey2021can} for the density evolution under the action of completely deterministic delayed dynamics \cite[Eqn. 15]{mackey2021can}.  Both of these results for the Liouville and Fokker-Planck like evolution equations are equivalent to those derived by others, e.g. \citet{guillouzic99}, \citet{loos2021stochastic} although our method of derivation deviates from theirs.\footnote{
As we have noted previously (\citet{losson2020density,mackey2021can}) utilizing and studying these evolution equations is dependent on having a well developed theory of integration for functionals which is generally lacking.  However, there is one situation in which we do have a very well developed integration theory and that revolves around the Wiener measure, and we have utilized this body of knowledge both in \citet{mackey2021can} and here to study linear delayed dynamics.  We have been able to examine the density evolution behaviour for both delayed and stochastic delayed linear systems.}

The interesting finding is that the presence of delay in a linear dynamics   can destroy monotonicity and lead to an oscillatory behaviour of the Gibbs' and conditional entropy, and thus $H_{NE}$.  If stochastic perturbations are simultaneously present in a linear system, then there may be an oscillatory approach of the entropies to a maximum.  The extent to which this behaviour will persist or disappear in nonlinear systems remains a problem for the future.

The next obvious extension will be to look at nonlinear delayed and stochastic delayed systems though it is not obvious to us how to proceed with this programme.

\vskip 0.5cm

\begin{filecontents}[overwrite]{\jobname.bib}

@ARTICLE{abbond99,
  AUTHOR =       "A. Abbondandolo",
  TITLE =        "An {H}-theorem for a class of {M}arkov processes",
  JOURNAL =      "Stoch. Anal. Applic.",
  YEAR =         "1999",
  volume =       "17",
doi = {10.1080/07362999908809593},
  pages =        "131-136"
}

@article {arnold01,
    AUTHOR = {Arnold, Anton and Markowich, Peter and Toscani, Giuseppe and
              Unterreiter, Andreas},
     TITLE = {On convex {S}obolev inequalities and the rate of convergence
              to equilibrium for {F}okker-{P}lanck type equations},
   JOURNAL = {Comm. Partial Differential Equations},
  FJOURNAL = {Communications in Partial Differential Equations},
    VOLUME = {26},
      YEAR = {2001},
DOI= {10.1081/PDE-100002246},
     PAGES = {43--100},
}

@ARTICLE{bag02,
  AUTHOR =       "B. C. Bag",
  TITLE =        "Nonequilibrium stochastic processes: {t}ime dependence of entropy flux and entropy production",
  JOURNAL =      "Phys. Rev. E",
  YEAR =         "2002",
  volume =       "66",
doi="10.1103/PhysRevE.66.026122",
  pages =        "026122-1-8"
}

@ARTICLE{bag02b,
  AUTHOR =       "B. C. Bag",
  TITLE =        "Upper bound for the time derivative of entropy for nonequilibrium stochastic processes",
  JOURNAL =      "Phys. Rev. E",
  YEAR =         "2002",
  volume =       "65",
doi="10.1103/PhysRevE.65.046118",
  pages =        "046118-1-6"
}

@ARTICLE{bag03,
  AUTHOR =       "B. C. Bag",
  TITLE =        "Information entropy production in non-{M}arkovian systems",
  JOURNAL =      "J. Chem. Phys.",
  YEAR =         "2003",
  volume =       "119",
doi="10.1063/1.1596411",
  pages =        "4988-4990"
}

@incollection {bakryemery85,
    AUTHOR = {Bakry, D. and {\'E}mery, M.},
     TITLE = {Diffusions hypercontractives},
 BOOKTITLE = {S\'eminaire de probabilit\'es, XIX, 1983/84},
    SERIES = {Lecture Notes in Math.},
    VOLUME = {1123},
doi={10.1007/BFb0075847},
     PAGES = {177--206},
 PUBLISHER = {Springer},
   ADDRESS = {Berlin},
      YEAR = {1985},
}

@ARTICLE{mazur01,
  AUTHOR =       "D. Bedeaux and P. Mazur",
  TITLE =        "Mesoscopic non-equilibrium thermodynamics for quantum systems",
  JOURNAL =      "Physica A",
  YEAR =         "2001",
  volume =       "298",
doi="10.1016/S0378-4371(01)00223-0",
  pages =        "81-100"
}

@Book{Boltzmann1872,
  author    = {Boltzmann, L.},
  publisher = {Aus der kk Hot-und Staatsdruckerei},
  title     = {Weitere {S}tudien {\"u}ber das {W}{\"a}rmegleichgewicht unter {G}asmolek{\"u}len},
  year      = {1872},
}

@Article{brillouin1950thermodynamics,
  author    = {Brillouin, L.},
  journal   = {American Scientist},
  title     = {Thermodynamics and information theory},
  year      = {1950},
  number    = {4},
  pages     = {594--599},
  volume    = {38},
  publisher = {JSTOR},
}

@ARTICLE{nicolis99,
  author = "D. Daems and G. Nicolis",
  TITLE =        "Entropy production and phase space volume contraction",
  JOURNAL =      "Phys. Rev. E.",
  YEAR =         "1999",
  volume =       "59",
doi="10.1103/PhysRevE.59.4000",
  pages =        "4000-4006"
}

@book{davies1977physics,
  title={The physics of time asymmetry},
  author={Davies, Paul Charles William},
  year={1977},
  publisher={Univ of California Press}
}

@BOOK{degroot84,
  AUTHOR =       "S. R. de Groot and P. Mazur",
  TITLE =        "Non-Equilibrium Thermodynamics",
  PUBLISHER =    "Dover",
  YEAR =         "1984",
  address = "New York"
}

@Book{dynkin1965markov,
  author    = {Dynkin, E. B.},
  publisher = {Academic Press Inc., Publishers, New York; Springer-Verlag, Berlin-G\"{o}ttingen-Heidelberg},
  title     = {Markov processes. {V}ols. {I}, {II}},
  year      = {1965},
}

@book{eddington2012new,
  title={New pathways in science: Messenger lectures (1934)},
  author={Eddington, Arthur},
  year={2012},
  publisher={Cambridge University Press}
}

@book{fer1977irreversibilite,
  title={L'irr{\'e}versibilit{\'e}, fondement de la stabilit{\'e} du monde physique},
  author={Fer, Francis},
  year={1977},
  publisher={Gauthier-Villars Paris}
}

@incollection{frigg2016field,
  title={A field guide to recent work on the foundations of statistical mechanics},
  author={Frigg, R.},
  booktitle={The Ashgate companion to contemporary philosophy of physics},
  pages={105--202},
  year={2016},
  publisher={Routledge}
}

@article{frigg2019statistical,
  title={Statistical mechanics: A tale of two theories},
  author={Frigg, R. and Werndl, C.},
  journal={The Monist},
  volume={102},
  number={4},
  pages={424--438},
  doi={10.1093/monist/onz018},
  year={2019},
  publisher={Oxford University Press}
}

@BOOK{gardinerhandbook,
  AUTHOR =       "C. W. Gardiner",
  TITLE =        "Handbook of Stochastic Methods",
  PUBLISHER =    "Springer Verlag",
  YEAR =         "1983",
  address =      "Berlin, Heidelberg",
  isbn =         "0-387-15607-0",
}

@article{gardner1967can,
  title={Can time go backward?},
  author={Gardner, Martin},
  journal={Scientific American},
  volume={216},
  number={1},
  pages={98--109},
  year={1967},
  publisher={JSTOR}
}

@BOOK{gibbs02,
  AUTHOR =       "J. W. Gibbs",
  TITLE =        "Elementary Principles in Statistical Mechanics",
  PUBLISHER =    "Dover",
  YEAR =         "1962",
  address =      "New York",
}

@article{gold1962arrow,
  title={The arrow of time},
  author={Gold, Thomas},
  journal={American Journal of Physics},
  volume={30},
  number={6},
  pages={403--410},
  doi={10.1119/1.1942052},
  year={1962},
  publisher={American Association of Physics Teachers}
}

@article{gold1969nature,
  title={The nature of time},
  author={Gold, Thomas and Schumacher, D. L.},
  journal={British Journal for the Philosophy of Science},
  volume={20},
  number={1},
  year={1969}
}

@ARTICLE{guillouzic99,
  AUTHOR =       {S. Guillouzic and I. L'Heureux and A. Longtin},
  TITLE =        {Small delay approximation of stochastic delay differential equations},
  JOURNAL =      {Phys. Rev. E},
  YEAR =         {1999},
  volume =       {61},
  doi={10.1103/PhysRevE.59.3970},
  pages =        {3970-3982},
}

@book {hale-lunel,
    AUTHOR = {Hale, Jack K. and Verduyn Lunel, Sjoerd M.},
     TITLE = {Introduction to functional-differential equations},
    SERIES = {Applied Mathematical Sciences},
    VOLUME = {99},
 PUBLISHER = {Springer-Verlag},
   ADDRESS = {New York},
      YEAR = {1993},
     PAGES = {x+447}
}

@book{halliwell1996physical,
  title={Physical origins of time asymmetry},
  author={Halliwell, Jonathan J. and P{\'e}rez-Mercader, Juan and Zurek, Wojciech Hubert},
  year={1996},
  publisher={Cambridge University Press}
}

@article{hayes1950roots,
  title={Roots of the transcendental equation associated with a certain difference-differential equation},
  author={Hayes, N. D.},
  journal = {J. London Math. Soc. (2)},
  fjournal={Journal of the London Mathematical Society},
  volume={1},
  number={3},
  pages={226--232},
  doi={10.1112/jlms/s1-25.3.226},
  year={1950},
  publisher={Wiley Online Library},
}

@book{hill2013statistical,
  title={Statistical {M}echanics: {P}rinciples and {S}elected {A}pplications},
  author={Hill, T. L.},
  year={2013},
  publisher={Courier Corporation}
}

@Article{Holubec2022,
  author    = {Holubec, V. and Ryabov, A. and Loos, S. A. M. and Kroy, K.},
  journal   = {New Journal of Physics},
  title     = {Equilibrium Stochastic Delay Processes},
  volume={24},
  doi={10.1088/1367-2630/ac4b91},
  year      = {2022},
  publisher = {IOP Publishing},
  pages={023021}
}

 @BOOK{horsthemke84,
  AUTHOR =       "W. Horsthemke and R. Lefever",
  TITLE =        "Noise Induced Transitions: Theory and Applications in Physics, Chemistry, and Biology",
  PUBLISHER =    "Springer-Verlag",
  YEAR =         "1984",
  address =      "Berlin, New York, Heidelberg"
}

@article {Huang,
    AUTHOR = {Huang, X. and R\"{o}ckner, M. and Wang, F.-Y.},
     TITLE = {Nonlinear {F}okker-{P}lanck equations for probability measures
              on path space and path-distribution dependent {SDE}s},
   JOURNAL = {Discrete Contin. Dyn. Syst.},
  FJOURNAL = {Discrete and Continuous Dynamical Systems. Series A},
    VOLUME = {39},
    doi={10.3934/dcds.2019125},
      YEAR = {2019},
    NUMBER = {6},
     PAGES = {3017--3035}
 }

 @ARTICLE{jaynes65,
  AUTHOR =       "E. Jaynes",
  TITLE =        "Gibbs vs {B}oltzmann entropies",
  JOURNAL =      "Amer. J. Phys.",
  YEAR =         "1965",
  volume =       "33",
  doi="10.1119/1.1971557",
  pages =        "391-398"
}

@article{jaynes1971violation,
  title={Violation of {B}oltzmann's {H} theorem in real gases},
  author={Jaynes, E. T.},
  journal={Physical Review A},
  volume={4},
  number={2},
  pages={747},
  doi={10.1103/PhysRevA.4.747},
  year={1971},
  publisher={APS}
}

@book{kittel2004elementary,
  title={Elementary {S}tatistical {P}hysics},
  author={Kittel, C.},
  year={2004},
  publisher={Courier Corporation}
}

@article{kuchler1992,
    AUTHOR = {K\"{u}chler, U. and Mensch, B.},
     TITLE = {Langevin's stochastic differential equation extended by a
              time-delayed term},
   JOURNAL = {Stochastics Stochastics Rep.},
  FJOURNAL = {Stochastics and Stochastics Reports},
    VOLUME = {40},
      YEAR = {1992},
      doi={10.1080/17442509208833780},
    NUMBER = {1-2},
     PAGES = {23--42},
 }

@Book{landsberg1985enigma,
  author    = {Landsberg, P. T.},
  publisher = {Hilger},
  title     = {The Enigma of Time},
  year      = {1985},
}

@Book{almcmbk94,
  author    = {A. Lasota and M. C. Mackey},
  publisher = {Springer-Verlag},
  title     = {Chaos, Fractals and Noise: Stochastic Aspects of Dynamics},
  year      = {1994},
    SERIES = {Applied Mathematical Sciences},
    VOLUME = {97},
  address   = {Berlin, New York, Heidelberg}
}

@ARTICLE{lebowitz93,
  AUTHOR =       "J. Lebowitz",
  TITLE =        "Macroscopic laws, microscopic dynamics, time's arrow and {B}oltzmann entropy",
  JOURNAL =      "Physica A",
  YEAR =         "1993",
  volume =       "194",
  pages =        "1-27",
  doi="10.1016/0378-4371(93)90336-3"
}

@Article{lebowitz99,
  author  = {J. L. Lebowitz},
  journal = {Physica A},
  title   = {Microscopic origins of irreversible macroscopic behavior},
  year    = {1999},
  pages   = {516-527},
  volume  = {263},
  doi={10.1016/S0378-4371(98)00514-7}
}

@ARTICLE{lebowitz99a,
  AUTHOR =       "J. L. Lebowitz",
  TITLE =        "Statistical mechanics: {A} selective review of two central issues",
  JOURNAL =      "Rev. Mod. Phys.",
  YEAR =         "1999",
  volume =       "71",
  pages =        "S346-S357",
  doi="10.1103/RevModPhys.71.S346"
}

@Book{loos2021stochastic,
  author    = {Loos, S. A. M.},
  publisher = {Springer Nature},
  title     = {Stochastic Systems with Time Delay: Probabilistic and Thermodynamic Descriptions of Non-Markovian Processes Far from Equilibrium},
  year      = {2021},
  doi={10.1007/978-3-030-80771-9}
}

@ARTICLE{loskot91,
  AUTHOR =       "K. Loskot and R. Rudnicki",
  TITLE =        "Relative entropy and stability of stochastic semigroups",
  JOURNAL =      "Ann. Pol. Math.",
  YEAR =         "1991",
  volume =       "52",
  pages =        "140-145",
  doi="10.4064/ap-53-2-139-145"
}

@Book{losson2020density,
  author    = {Losson, J. and Mackey, M. C. and Taylor, R. and Tyran-Kami{\'n}ska, M.}, 
  publisher = {Springer, New York},
  title     = {Density {E}volution {U}nder {D}elayed {D}ynamics: {A}n {O}pen {P}roblem},
  year      = {2020},
  series    = {Fields Institute Monographs},
  volume    = {38},
  doi={10.1007/978-1-0716-1072-5}
}

@Book{mackey2011time,
  author    = {Mackey, M. C.},
  publisher = {Springer Verlag},
  title     = {Time's Arrow: The Origins of Thermodynamic Behavior},
  year      = {2011},
}

@article{mackey1995solution,
  title={Solution moment stability in stochastic differential delay equations},
  author={Mackey, M. C. and Nechaeva, I. G.},
  journal={Physical Review E},
  volume={52},
  number={4},
  pages={3366},
  year={1995},
  publisher={APS},
  doi={10.1103/PhysRevE.52.3366}
}

@article{mackey2006noise,
  title={Noise and conditional entropy evolution},
  author={Mackey, M. C. and Tyran-Kami{\'n}ska, M.},
  journal={Physica A},
  volume={365},
  number={2},
  pages={360--382},
  year={2006},
  publisher={Elsevier},
  doi={10.1016/j.physa.2005.10.001}
}

@article{mackey2006temporal,
  title={Temporal behavior of the conditional and {G}ibbs’ entropies},
  author={Mackey, M. C. and Tyran-Kami{\'n}ska, M.},
  journal={J.~Stat. Phys.},
  volume={124},
  pages={1443--1470},
  year={2006},
  doi={10.1007/s10955-006-9181-0},
  publisher={Springer}
}

@Article{mackey2021can,
  author    = {Mackey, M. C. and Tyran-Kami{\'n}ska, M.},
  journal   = {Chaos},
  title     = {How can we describe density evolution under delayed dynamics?},
  year      = {2021},
  number    = {4},
  pages     = {043114},
  volume    = {31},
  publisher = {AIP Publishing LLC},
  doi={10.1063/5.0038310}
}

@book{mayer1940statistical,
  title={Statistical Mechanics},
  author={Mayer, J. E. and Mayer, M. G.},
  volume={28},
  year={1940},
  publisher={John Wiley \& Sons New York}
}

@ARTICLE{mazur98b,
  AUTHOR =       "P. Mazur",
  TITLE =        "Fluctuations and non-equilibrium thermodynamics",
  JOURNAL =      "Physica A",
  YEAR =         "1998",
  volume =       "261",
  pages =        "451-457",
  doi="10.1016/S0378-4371(98)00353-7"
}

@Book{mohammed84,
  author    = {S-E A. Mohammed},
  publisher = {Pitman},
  title     = {Stochastic Functional Differential Equations},
  year      = {1984},
  address   = {Boston}
}

@Article{nicolis98,
  author  = {G. Nicolis and D. Daems},
  journal = {Chaos},
  title   = {Probabilistic and thermodynamic aspects of dynamical systems},
  year    = {1998},
  pages   = {311-320},
  volume  = {8},
  doi={10.1063/1.166313}
}

@Book{Penrose2005,
  author    = {O. Penrose},
  publisher = {Dover},
  title     = {Foundations of Statistical Mechanics},
  year      = {2005},
  address   = {Mineola, New York},
  edition   = {Revised},
}

@Article{mazur94,
  author  = {A. P{'e}rez-Madrid and J. M. Rub{'i} and P. Mazur},
  journal = {Physica A},
  title   = {Brownian motion in the presence of a temperature gradient},
  year    = {1994},
  pages   = {231-238},
  volume  = {212},
  doi={10.1016/0378-4371(94)90329-8}
}

@article {pichorrudnicki00,
    AUTHOR = {Pich{\'o}r, K. and Rudnicki, R.},
     TITLE = {Continuous {M}arkov semigroups and stability of transport
              equations},
   JOURNAL = {J. Math. Anal. Appl.},
     VOLUME = {249},
      YEAR = {2000},
       PAGES = {668--685},
       doi={10.1006/jmaa.2000.6968}
    }

@ARTICLE{qian01,
  AUTHOR =       "H. Qian",
  TITLE =        "Relative entropy: {F}ree energy associated with equilibrium fluctuations and nonequilibrium deviations",
  JOURNAL =      "Phys. Rev. E.",
  YEAR =         "2001",
  volume =       "63",
  pages =        "042103-1-5",
  doi="10.1103/PhysRevE.63.042103"
}

@ARTICLE{qian02a,
  AUTHOR =       "H. Qian",
  TITLE =        "Equations for stochastic macromolecular mechanics of single proteins: {E}quilibriuim fluctuations,
  transient kinetics and nonequilibrium steady-state",
  JOURNAL =      "J. Phys. Chem.",
  YEAR =         "2002",
  volume =       "106",
  pages =        "2065-2073",
  doi="10.1021/jp013143w"
}

@ARTICLE{qian02,
  AUTHOR =       "H. Qian and M. Qian and X. Tang",
  TITLE =        "Thermodynamics of the general diffusion process: {T}ime-reversibility and entropy production",
  JOURNAL =      "J. Stat. Phys.",
  YEAR =         "2002",
  volume =       "107",
  pages =        "1129-1141",
  doi="10.1023/A:1015109708454"
}

@BOOK{reichenbach57,
  AUTHOR =       "H. Reichenbach",
  TITLE =        "The Direction of Time",
  PUBLISHER =    "California University Press",
  YEAR =         "1957",
  address =      "Berekeley",
}

@BOOK{reichlbk,
  AUTHOR =       "L. E. Reichl",
   TITLE =        "A Modern Course in Statistical Physics",
  PUBLISHER =    "University of Texas Press",
  YEAR =         "1980",
  address =      "Austin"
}

@BOOK{risken84,
  AUTHOR =       "H. Risken",
  TITLE =        "The Fokker-Planck Equation",
  PUBLISHER =    "Springer-Verlag",
  YEAR =         "1984",
  address =      "Berlin, New York, Heidelberg",
  doi="10.1007/978-3-642-61544-3"
}

@ARTICLE{mazur98a,
  AUTHOR =       "J. M. Rub{'i} and P. Mazur",
  TITLE =        "Simultaneous {B}rownian motion of {N} particles in a temperature gradient",
  JOURNAL =      "Physica A",
  YEAR =         "1998",
  volume =       "250",
  pages =        "253-264",
  doi="10.1016/S0378-4371(97)00463-9"
}

@ARTICLE{mazur00,
  AUTHOR =       "J. M. Rub{'i} and P. Mazur",
  TITLE =        "Nonequilibrium thermodynamics and hydrodynamic fluctuations",
  JOURNAL =      "Physica A",
  YEAR =         "2000",
  volume =       "276",
  pages =        "477-488",
  doi="10.1016/S0378-4371(99)00452-5"
}

@ARTICLE{rubi01,
  AUTHOR =       "J. M. Rub{'i} and A. P{'e}rez-Madrid",
  TITLE =        "Mesocopic non-equilibrium thermodynamics approach to the dynamics of polymers",
  JOURNAL =      "Physica A",
  YEAR =         "2001",
  volume =       "298",
  pages =        "177-186",
  doi="10.1016/S0378-4371(01)00217-5"
}

@ARTICLE{ruelle96,
  AUTHOR =       "D. Ruelle",
  TITLE =        "Positivity of entropy production in nonequilibrium statistical mechanics",
  JOURNAL =      "J. Stat. Phys.",
  YEAR =         "1996",
  volume =       "85",
  pages =        "1-22",
  doi="10.1007/BF02175553"
}

@ARTICLE{ruelle97,
  AUTHOR =       "D. Ruelle",
  TITLE =        "Entropy production in nonequilibrium statistical mechanics",
  JOURNAL =      "Commun. Math. Phys.",
  YEAR =         "1997",
  volume =       "189",
  pages =        "365-371",
  doi="10.1007/s002200050207"
}

@ARTICLE{ruelle03,
  AUTHOR =       "D. Ruelle",
  TITLE =        "Extending the definition of entropy to nonequilibriuim steady states",
  JOURNAL =      "Proc. Nat. Acad. Sci.",
  YEAR =         "2003",
  volume =       "100",
  pages =        "3054-3058",
  doi="10.1073/pnas.0630567100"
}

@ARTICLE{ruelle04,
  AUTHOR =       "D. Ruelle",
  TITLE =        "Conversations on nonequilibrium physics with an extraterrestrial",
  JOURNAL =      "Phys. Today",
  YEAR =         "2004",
  volume =       "189",
month="May",
  pages =        "48-53",
  doi="10.1063/1.1768674"
}

@BOOK{sachs87,
  AUTHOR =       "R. G. Sachs",
  TITLE =        "The Physics of Time Reversal",
  PUBLISHER =    "University of
Chicago Press",
  YEAR =         "1987",
  address =      "Chicago",
}

@Book{savitt1998time,
  author    = {Savitt, S. F.},
  title     = {Time's {A}rrows {T}oday: {R}ecent {P}hysical and {P}hilosophical {W}ork on the {D}irection of {T}ime},
  year      = {1998},
  publisher = {Cambridge University Press}
}

@book{schulman97,
author="L. S. Schulman",
title="Time's Arrows and Quantum Measurement",
year="1997",
publisher="Cambridge University Press",
address="Cambridge"
}

@Article{schulman2010we,
  author    = {Schulman, L. S.},
  journal   = {Physica E},
  title     = {We know why coffee cools},
  year      = {2010},
  number    = {3},
  pages     = {269--272},
  volume    = {42},
  publisher = {Elsevier},
  doi={10.1016/j.physe.2009.06.045}
}

@BOOK{sklar93,
  AUTHOR =       "L. Sklar",
  TITLE =        "Introduction to Physics and Chance. Philospohical Issues
  in the Foundations of Statistical Mechanics",
  PUBLISHER =    "Cambridge University Press",
  YEAR =         "1993",
  address =      "Cambridge"
}

@book{ter1995elements,
  title={Elements of Statistical Mechanics},
  author={Ter Haar, D.},
  year={1995},
  publisher={Elsevier}
}

@ARTICLE{toscanivillani,
  AUTHOR =       "G. Toscani and C. Villani",
  TITLE =        "On the trend to equilibrium for some dissipative systems with slowly increasing a priori bounds",
  JOURNAL =      "J. Stat. Phys.",
  YEAR =         "2000",
  volume =       "98",
  pages =        "1279-1309",
  doi="10.1023/A:1018623930325"
}

@UNPUBLISHED{uffink06,
  author =       {J. Uffink},
  title =        {Compendium of the foundations of classical statistical physics},
  note =         {PhilSci Archive},
  year =         {2006},
  source =       {http://philsci-archive.pitt.edu/2691/},
}

@InCollection{uffink2022sep-statphys-Boltzmann,
	author       =	{Uffink, J.},
	title        =	{{Boltzmann’s Work in Statistical Physics}},
	booktitle    =	{The {Stanford} Encyclopedia of Philosophy},
	editor       =	{E.N. Zalta},
	howpublished =	{\url{https://plato.stanford.edu/archives/sum2022/entries/statphys-Boltzmann/}},
	year         =	{2022},
	edition      =	{{S}ummer 2022},
	publisher    =	{Metaphysics Research Lab, Stanford University}
}

@Article{VanKampen1981,
  author    = {Van Kampen, N. G.},
  journal   = {J. Stat. Phys.},
  title     = {It{\^o} versus {S}tratonovich},
  year      = {1981},
  number    = {1},
  pages     = {175--187},
  volume    = {24},
  publisher = {Springer},
  doi={10.1007/BF01007642}
}

@book{van1992stochastic,
  title={Stochastic {P}rocesses in {P}hysics and {C}hemistry},
  author={Van Kampen, N. G.},
  volume={1},
  year={1992},
  publisher={Elsevier}
}

@Book{zeh1992physical,
  author= {Zeh, H. D.},
  publisher = {Springer},
  title= {The Physical Basis for the Direction of Time},
  series={Second edition},
  year= {1992},
}
\end{filecontents}

\centerline{\bf\large References}
\bibliographystyle{abbrvnat}
\bibliography{\jobname}
\bigskip

\Koniec
\end{document}